\pdfoutput=1

\documentclass[11pt]{article}

\usepackage[]{acl}

\usepackage{times}
\usepackage{latexsym}

\usepackage[T1]{fontenc}

\usepackage[utf8]{inputenc}

\usepackage{microtype}

\usepackage{inconsolata}

\usepackage{tikz}
\usetikzlibrary{shapes.arrows}

\usepackage[listings]{tcolorbox}
\usepackage{amsmath}
\usepackage{amsfonts}       
\usepackage{multirow}
\usepackage{tabularx}
\usepackage{booktabs}       
\usepackage{graphicx}
\usepackage{subcaption}
\usepackage{multirow}
\usepackage{enumitem}
\usepackage{url}
\usepackage{adjustbox}
\newcommand{\ie}{\emph{i.e., }}
\newcommand{\eg}{\emph{e.g., }}

\newcommand{\etc}{\emph{etc.}}

\newcommand{\cf}{\emph{cf. }}

\newcommand{\myarrow}[1][]{\mathrel{\tikz{
    \node[fill, minimum width=1.2ex, minimum height=.9em, inner sep=0pt, single arrow, single arrow head extend=2pt, single arrow tip angle=60](A){\raisebox{3.5pt}[0pt][0pt]{$\,\scriptstyle #1\ $}}; \path([xshift=-.4pt]A.west)--(A.east);}}}

\newcommand{\shiyr}[1]{\textcolor{black}{#1}}

\definecolor{red_orange}{HTML}{FA5722}
\definecolor{sky_blue}{HTML}{6D9CF5}
\definecolor{mygreen}{RGB}{2, 142, 2}
\usepackage[symbol]{footmisc}

\title{ReactXT: Understanding Molecular “Reaction-ship” via Reaction-Contextualized Molecule-Text Pretraining}

\author{Zhiyuan Liu$^{1 *}$ \quad Yaorui Shi$^{2 *}$ \quad An Zhang$^1$\quad Sihang Li$^2$ \\ \textbf{Enzhi Zhang$^3$ \quad Xiang Wang$^{2 \dag}$ \quad Kenji Kawaguchi$^1$ \quad Tat-Seng Chua$^1$}\\
$^1$National University of Singapore \\
$^2$University of Science and Technology of China \quad 
$^3$Hokkaido University\\
\texttt{\{acharkq,yaoruishi,an.zhang3.14,sihang0520,xiangwang1223\}@gmail.com}\\ \texttt{enzhi.zhang.n6@elms.hokudai.ac.jp}, \texttt{\{kenji,chuats\}@comp.nus.edu.sg}\\}

\begin{document}
\maketitle
\begin{abstract}
Molecule-text modeling, which aims to facilitate molecule-relevant tasks with a textual interface and textual knowledge, is an emerging research direction. Beyond single molecules, studying reaction-text modeling holds promise for helping the synthesis of new materials and drugs. However, previous works mostly neglect reaction-text modeling: they primarily focus on modeling individual molecule-text pairs or learning chemical reactions without texts in context. Additionally, one key task of reaction-text modeling -- experimental procedure prediction -- is less explored due to the absence of an open-source dataset. The task is to predict step-by-step actions of conducting chemical experiments and is crucial to automating chemical synthesis. To resolve the challenges above, we propose a new pretraining method, \textbf{ReactXT}, for reaction-text modeling, and a new dataset, \textbf{OpenExp}, for experimental procedure prediction. Specifically, ReactXT features three types of input contexts to incrementally pretrain LMs. Each of the three input contexts corresponds to a pretraining task to improve the text-based understanding of either reactions or single molecules. 
ReactXT demonstrates consistent improvements in experimental procedure prediction and molecule captioning and offers competitive results in retrosynthesis. Our code is available at \url{https://github.com/syr-cn/ReactXT}.
\end{abstract}


\begin{figure}
\end{figure}

\section{Introduction}
\footnotetext[1]{Equal contribution.}
\footnotetext[2]{Corresponding author. Xiang Wang is also affiliated with Institute of Dataspace, Hefei Comprehensive National Science Center.}
Multi-modal large language models (LMs) have recently attracted extensive research attention. Remarkably, in the vision-language domain, LMs enhanced with visual encoders show impressive results in visual question-answering and image captioning~\cite{llava,BLIP2}. Inspired by their successes, molecule-text modeling (MTM) becomes an emerging research field~\cite{MolCA, KVPLM, MoMu}, aiming to build the natural language interface for molecular tasks, including text-guided molecule generation, molecule captioning, and molecule-text retrieval~\cite{MolT5,MoleculeSTM}.

\begin{figure}
\end{figure}



\begin{figure*}[ht]
\centering \small
\includegraphics[width=0.95\textwidth]{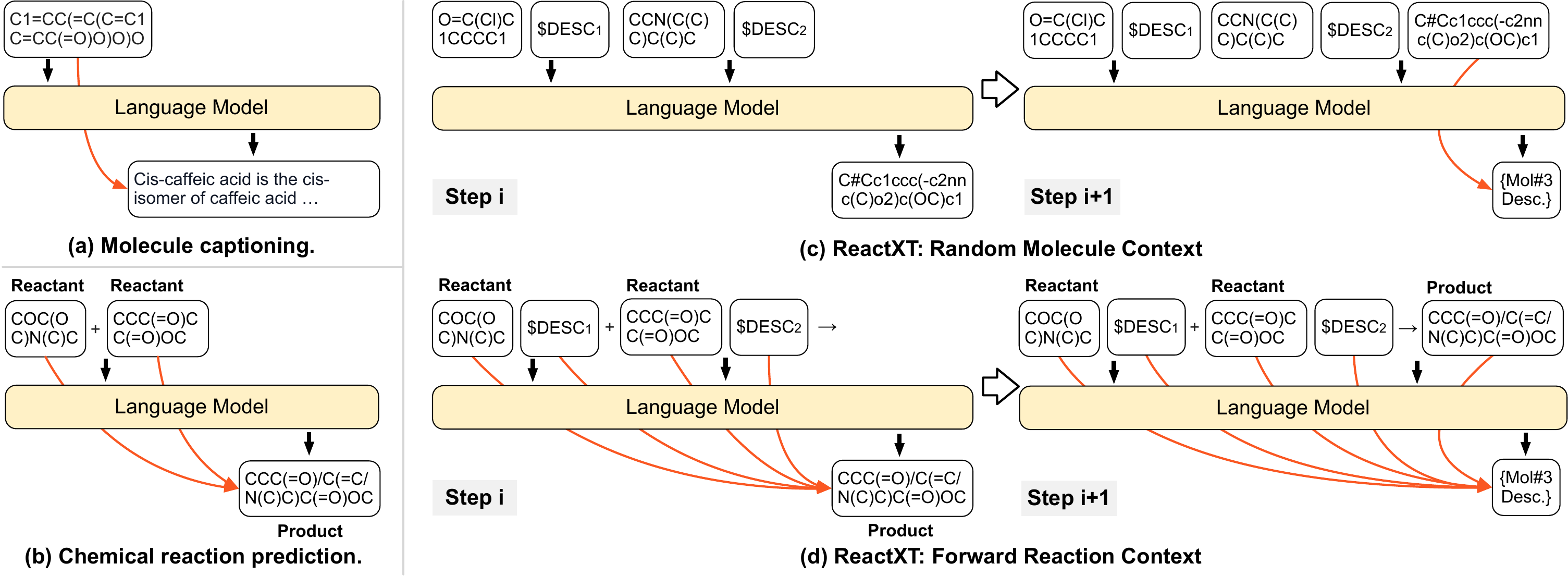}
\caption{Comparison of molecule-text generative modeling methods. \textcolor{red_orange}{Orange arrows $\myarrow$} 
denote the chemical relations for generation. 2D graph embeddings~\cite{MolCA} are omitted here for simplicity, but are added in the final framework for improved performance. \texttt{\$}$\mathtt{DESC_j}$ denotes the description of the $j$-th molecule. The chemical reaction in Figures (b) and (d) is: COC(OC)N(C)C + CCC(=O)CC(=O)OC $\rightarrow$ CCC(=O)/C(=C/N(C)C)C(=O)OC.
}
\label{fig:comparison}
\end{figure*}

\begin{figure*}[ht]
\centering \small
\includegraphics[width=0.95\textwidth]{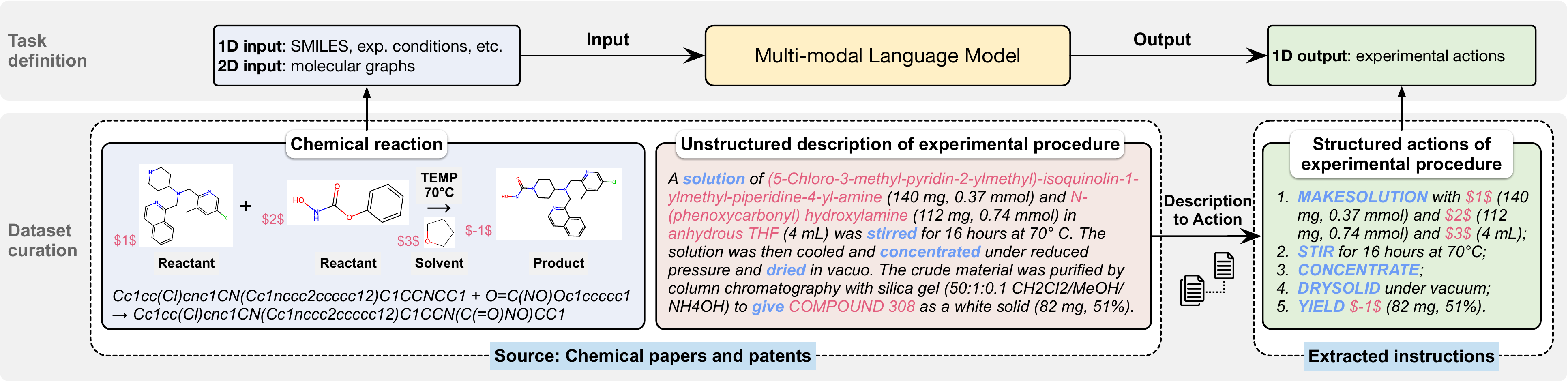}
\caption{Illustration of the experimental procedure prediction task and its dataset curation process. We employ the actions defined by~\cite{smiles2actions} and the description to action model from~\cite{TextChemT5}.}
\label{fig:task}
\vspace{-3mm}
\end{figure*}

Building upon these MTM works, we study reaction-text modeling (RTM), aiming to improve LMs' performance on reaction-relevant tasks. Chemical reactions, involving the transformation of reactants into products, are fundamental to advancing drug discovery and material science~\cite{schwaller2022machine}.
Revisiting prior works, we identify key research gaps in both the learning paradigm and the evaluation benchmark for RTM:
\begin{itemize}[leftmargin=*]
\item \textbf{Learning Paradigm.} Most prior works either focus on generating the textual description of a single molecule (\cf Figure~\ref{fig:comparison}a)~\cite{MolCA, MolT5, MoMu}, or apply LMs for chemical reaction prediction without including the textual descriptions of molecules/reactions in context (\cf Figure~\ref{fig:comparison}b)~\cite{TextChemT5,MolInstructions,RegressionTransformer}. Such methods overlook the potential knowledge in textual descriptions to improve performance. Pioneer works~\cite{relm,GPTChemBench} include labels of molecular roles and experimental conditions when prompting ChatGPT, but achieve suboptimal performances for being limited to prompt engineering.
\item \textbf{Evaluation Benchmark.} An open-source dataset for experimental procedure prediction is notably missing. 
As illustrated in Figure~\ref{fig:task}, experimental procedure prediction aims to deduce the step-by-step actions for experimental execution through interpreting chemical reactions~\cite{smiles2actions}, which has a significant value for automating chemical synthesis processes~\cite{paragraph2actions,description2instructions}. This task aligns well with our focus on RTM, requiring an understanding of chemical reactions and a textual interface to articulate experimental steps. Unfortunately, the absence of public datasets hinders further research and development in this area.
\end{itemize}


Addressing the identified research gaps, we propose \underline{React}ion-Conte\underline{xt}ualized Molecule-Text Pretraining (\textbf{ReactXT}), aiming to improve the text-based understanding of chemical reactions and molecules. Further, we construct an \underline{open}-source dataset for \underline{exp}erimental procedure prediction (\textbf{OpenExp}), serving as a key benchmark to evaluate RTM methods. Below, we elaborate on their details.

\textbf{ReactXT} aims to improve the learning paradigm of RTM by introducing three types of input contexts, each of which corresponds to a pretraining task to improve LMs' understanding of chemical reactions or individual molecules. As Figure~\ref{fig:comparison}d depicts, the forward reaction context is crafted to learn the chemical connections among molecules involved in the same reaction. These connections are grounded on chemical reaction principles, such as the conservation laws~\cite{atkins2007chemical}. Building on this molecular interplay, we hypothesize that understanding other molecules in the same reaction and their descriptions can help predict the current molecule and its textual description. ReactXT encourages LMs to harness these inter-molecule relationships to improve their ability to generate molecular descriptions in reactions and, in turn, deepen their understanding of chemical reaction principles. Further, a backward reaction context is introduced to support retrosynthesis tasks (\cf Section~\ref{sec:input_context}). Finally, as Figure~\ref{fig:comparison}c illustrates, ReactXT includes the random molecule context, cultivating the LMs' understanding of individual molecules outside their reactions.

\textbf{OpenExp} features $274,439$ pairs of chemical reactions and their corresponding step-by-step instructions of experimental procedures. This dataset, compiled from the USPTO-Applications~\cite{USPTO-Applications} and ORD~\cite{OpenORD} databases, will be released under the CC-BY-SA license. To ensure data quality, we have conducted careful data preprocessing. Further, we invite human experts to evaluate the dataset quality. Out of 100 randomly chosen samples, 50 samples could be directly used without any human intervention, and 90 samples required only minor modifications for experimental execution (\cf Figure \ref{fig:human_evaluation}).


Our contributions can be summarized as follows:
\begin{itemize}[leftmargin=*]
    \item We propose ReactXT, a method that incorporates three types of input contexts to incrementally pretrain an LM. These contexts are tailored to enhance LMs' understanding of chemical reactions and individual molecules.
    \item We curate an open-source experimental procedure prediction dataset OpenExp, a new benchmark for automating chemical synthesis research.
    \item ReactXT achieves state-of-the-art performances for experimental procedure prediction on the OpenExp dataset, highlighting its superior RTM ability. It also outperforms baselines by 3.2\% for molecule captioning on the PubChem324k dataset. ReactXT has competitive performances for retrosynthesis, and we are refining it to surpass the current state-of-the-art method.
\end{itemize}

\section{Related Works}
\label{rel}

\textbf{Molecule-Text Modeling (MTM).} 
MTM aims to jointly model molecules and texts to address text-related molecular tasks~\cite{MolT5, Text2Mol}. Molecules can be represented by 1D sequences of SMILES~\cite{SMILES} and SELFIES~\cite{selfies}, making it feasible to pretrain unified LMs on mixed 1D sequences of texts and molecules~\cite{Galactica, MolT5, chemberta, KVPLM}. Further, these LMs can be aligned to human preference via instruction tuning~\cite{TextChemT5,MolInstructions}. In parallel to 1D LMs, multi-modal methods are also studied, using graph neural networks (GNNs)~\cite{pretrain_gnn, simsgt} to encode 2D molecular graphs. Notably, CLIP-style~\cite{CLIP} cross-modal contrastive learning and BLIP2-style~\cite{BLIP2} cross-modal projector are both investigated to facilitate molecule-text retrieval~\cite{MoMu, MoleculeSTM}, and molecule-to-text generation~\cite{MolCA, 3dmolm}, respectively. \shiyr{Recently, MolTC~\cite{moltc} is also proposed to model molecular interactions using chain of thoughts.} However, prior works mainly focus on individual molecules rather than chemical reactions. To bridge the gap, ReactXT explores reaction-text modeling, facilitating reaction-relevant tasks with a text interface and textual knowledge.

\textbf{Experimental Procedure Prediction.}
Synthesizing complex compounds requires detailed planning of synthetic pathways and intermediate steps, a process that is both labor-intensive and complex. Machine learning (ML) can potentially automate the process by predicting experimental procedures. Prior works have explored predicting reaction conditions (\eg catalyst and solvent)~\cite{gao2018using} and sequences of synthesis steps~\cite{smiles2actions} by reading chemical reactions. Given known experimental procedures, ML is also explored to empower chemical lab robots~\cite{MobileChemist}, and automated lab pipelines~\cite{coley2019robotic, nicolaou2020context}. Notably, tool-augmented GPT4~\cite{GPT4} is explored to plan and execute known chemical experiments~\cite{auto_research}. Unlike prior works, our OpenExp dataset is the first open-source dataset to facilitate the procedure prediction of unseen chemical experiments.

\textbf{Retrosynthesis and Chemical Reaction Prediction.}
Given a chemical reaction, retrosynthesis is to predict reactants from products and reaction prediction is to predict products from reactants~\cite{schwaller2022machine}. They can be formalized as sequence-to-sequence translation represented by SMILES strings~\cite{seq2seq, chemformer, r-smiles, AT, RetroTRAE}. 
Concurrently, 2D molecular graphs are explored for reaction prediction: selection-based methods focus on classifying the most suitable reaction templates \cite{localretro, GLN}; and graph-based generative models directly synthesize target molecules \cite{g2g, megan, RetroXpert}. However, the methods above leverage only reactions without texts. While notably two pioneer works apply ChatGPT for reaction prediction~\cite{relm,chemcrow}, their performances are limited to exploring only prompt engineering.

\begin{figure*}[ht]
\centering \small
\includegraphics[width=0.95\textwidth]{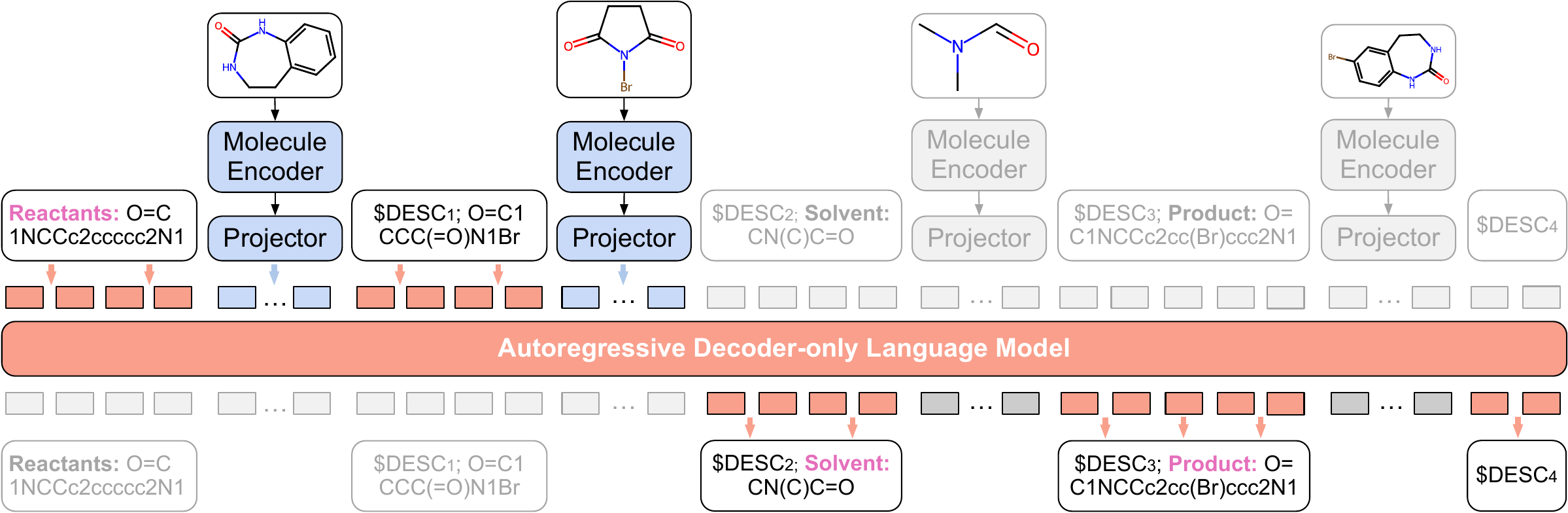}
\caption{Illustration of Reaction-Contextualized Molecule-Text Pretraining. Example uses forward reaction context.}
\vspace{-3mm}
\label{fig:framework}
\end{figure*}

\begin{table*}[t]
\scriptsize
\centering
\setlength{\tabcolsep}{2pt}
\scalebox{0.98}{
\begin{tabular}[t]{ll} \toprule
\textbf{Context Type}  & \textbf{Prompt Template} \\ \midrule
Forward reaction & \texttt{$\text{Reactants: } \underbrace{\texttt{\$}\mathtt{ SMI_1} \texttt{ <}\mathtt{Mol_1}\texttt{> } \texttt{\$}\mathtt{DESC_1;}}_{\times n \text{: Number of reactants}} \text{ Solvent: } \texttt{\$}\mathtt{ SMI_{n+1}} \texttt{ <}\mathtt{Mol_{n+1}}\texttt{> } \texttt{\$}\mathtt{DESC_{n+1};} \text{ Product: } \texttt{\$}\mathtt{ SMI_{n+2}} \texttt{ <}\mathtt{Mol_{n+2}}\texttt{> } \texttt{\$}\mathtt{DESC_{n+2}} \texttt{<STOP>}$} \\
Backward reaction & \texttt{$ \text{Product: } \texttt{\$}\mathtt{ SMI_1} \texttt{ <}\mathtt{Mol_1}\texttt{> } \texttt{\$}\mathtt{DESC_1;} \text{ Solvent: } \texttt{\$}\mathtt{ SMI_2} \texttt{ <}\mathtt{Mol_2}\texttt{> } \texttt{\$}\mathtt{DESC_2;} \text{ Reactants: } \underbrace{\texttt{\$}\mathtt{ SMI_3} \texttt{ <}\mathtt{Mol_3}\texttt{> } \texttt{\$}\mathtt{DESC_3}}_{\times n \text{: Number of reactants}} \texttt{<STOP>}$} \\
Random molecule  & \texttt{$ \texttt{\$}\mathtt{ SMI_1} \texttt{ <}\mathtt{Mol_1}\texttt{> } \texttt{\$}\mathtt{DESC_1;} \texttt{ \$}\mathtt{SMI_2} \texttt{ <}\mathtt{Mol_2}\texttt{> } \texttt{\$}\mathtt{DESC_2;} \texttt{ \$}\mathtt{SMI_3} \texttt{ <}\mathtt{Mol_3}\texttt{> } \texttt{\$}\mathtt{DESC_3;} \texttt{ \$}\mathtt{SMI_4} \texttt{ <}\mathtt{Mol_4}\texttt{> } \texttt{\$}\mathtt{DESC_4} \text{<STOP>}$} \\ \bottomrule
\end{tabular}
}
\caption{Prompt templates for creating input contexts. $\texttt{<}\mathtt{Mol_i}\texttt{>}$ is the placeholder for the 2D graph embedding of the i-th molecule; \texttt{\$$\mathtt{SMI_i}$} and \texttt{\$$\mathtt{DESC_i}$} is the SMILES and textual description for the i-th molecule, respectively.}
\vspace{-3mm}
\label{tab:prompt}
\end{table*}

\begin{table*}[h!]
\scriptsize
\centering
\setlength{\tabcolsep}{2pt}
\scalebox{0.98}{
\begin{tabular}[t]{p{15.6cm}} \toprule
\textbf{[Abstract]} The invention relates to indole acetic acid compounds which function as antagonists of the CRTH2 receptor. The invention also relates to the use of these compounds to inhibit the binding of prostaglandin D2 and its metabolites or certain thromboxane metabolites to the CRTH2 receptor and to treat disorders responsive to such inhibition. \textbf{[Properties]} Molecular Weight: 547.60; XLogP3: 6.10; Hydrogen Bond Donor Count: 0; Hydrogen Bond Acceptor Count: 7; Rotatable Bond Count: 8; Exact Mass: 547.19; Monoisotopic Mass: 547.19; Topological Polar Surface Area: 89.40; Heavy Atom Count: 39; Formal Charge: 0; Complexity: 1020; Isotope Atom Count: 0; Defined Atom Stereocenter Count: 0; Undefined Atom Stereocenter Count: 0; Defined Bond Stereocenter Count: 0; Undefined Bond Stereocenter Count: 0; Covalently-Bonded Unit Count: 1; Compound Is Canonicalized: Yes. \\ \bottomrule
\end{tabular}
}
\caption{Molecule description example, including the patent abstract and the computed/experimental properties. The described molecule is Cc1c(C2=NN(CCc3ccccc3)S(=O)(=O)c3ccccc32)c2cc(F)ccc2n1CC(=O)OC(C)(C)C.}
\vspace{-3mm}
\label{tab:example}
\end{table*}

\section{ReactXT: Reaction-Contextualized Molecule-Text Pretraining}
\label{sec:pretrain}
ReactXT consists of two key components: 1) the method of creating input contexts to incrementally pretrain an LM, and 2) a balanced sampling strategy for the reaction contexts. We begin by introducing our multi-modal LM backbone, then proceed to elaborate on ReactXT's two components.

\textbf{Multi-Modal Language Model Backbone.} Molecules can be represented by their 1D SMILES or 2D molecular graphs~\cite{wells2012structural}. We employ MolCA~\cite{MolCA} as our primary LM backbone to effectively harness both the 1D and 2D molecular modalities. Specifically, MolCA incorporates a GNN encoder~\cite{GraphCL} for encoding 2D molecular graphs. This GNN's output then is mapped to an LM's (\ie Galactica; \citet{Galactica}) input space via a cross-modal projector, thereby enabling the LM to perceive 2D molecular graphs. Both the cross-modal projector and the GNN have been pretrained for molecule-text alignment~\cite{BLIP2}. MolCA shows promising performances when finetuned for molecule captioning and IUPAC name prediction.



\subsection{Creating Input Contexts}
\label{sec:input_context}
Addressing the core challenges of LMs hinges on the careful selection of the input data. As shown in Table~\ref{tab:prompt}, ReactXT incorporates three types of input contexts to incrementally pretrain LMs: forward reaction context, backward reaction context, and random molecule context. These contexts are tailored for a text-based understanding of chemical reactions and individual molecules:
\begin{itemize}[leftmargin=*]
\item \textbf{Forward Reaction Context.} As Figure~\ref{fig:framework} illustrates, the forward reaction context labels molecules according to their roles -- \texttt{Reactant}, \texttt{Catalyst}, \texttt{Solvent}, and \texttt{Product} -- in the reaction, and arranges them in this specific sequential order. Note, not every reaction has a \texttt{Catalyst} or \texttt{Solvent}. For each molecule, we append its 2D molecular graph embeddings (\eg $\texttt{<}\mathtt{Mol_1}\texttt{>}$; \citet{MolCA}) after its SMILES to enhance the LM's understanding of molecular structures; and append molecular descriptions (\eg \texttt{\$}$\mathtt{DESC_1}$) following the 2D molecular graph embeddings to align molecules with texts.
\item \textbf{Backward Reaction Context.} Similar to the forward context but with the order of molecular roles reversed, this context aims to combat the Reversal Curse~\cite{ReversalCurse} of LMs: LMs trained on ``A is B'' fail to generalize to ``B is A''. The reversal generalization is crucial because downstream applications include backward retrosynthesis~\cite{schwaller2022machine}.
\item \textbf{Random Molecule Context.} Introduced to ensure LMs retain the capability to describe individual molecules outside chemical reactions.
\end{itemize}

\textbf{Context Length.} In each input context, we use up to $k$ molecules and their descriptions, where $k$ is a hyperparameter. For reactions with over $k$ molecules, we apply weighted molecule sampling, as explained in Section~\ref{sec:balance_samp}.


\textbf{Molecule Descriptions.} One crucial component of the input contexts is the molecule description, whose quality and comprehensiveness are vital for molecule-text alignment. We collect molecular descriptions and properties from multiple sources, encompassing three types of content:
\begin{itemize}[leftmargin=*]
    \item \textbf{Molecule Patent Abstracts.} We source patent abstracts from PubChem's Patent View\footnote{\url{https://pubchem.ncbi.nlm.nih.gov/docs/patents}}. These abstracts typically describe molecular structures, properties, or applications, but may also include irrelevant information if the molecule is merely mentioned in passing rather than being the central subject. Despite the noise, patent abstracts are indispensable for RTM: they cover ${\sim}95\%$ molecules in our employed reaction databases~\cite{USPTO-Applications, OpenORD}. In contrast, the molecule-text datasets~\cite{MoleculeSTM, MolCA} derived from PubChem's description section only cover ${\sim} 1\%$ of these molecules.
    \item \textbf{Computed and Experimental Properties.} We retrieve these numerical properties from PubChem, aiming to enhance the understanding of molecular structures through predictive learning. Certain properties are also helpful for reaction prediction. For example, knowing the solubility helps determine concentrations when preparing solutions; the knowledge of melting and boiling points helps identify the states of matter at given temperatures. Table~\ref{tab:example} shows an example of a patent abstract and computed/experimental properties. Table~\ref{tab:molecule_properties} includes detailed statistics of our collected molecule properties. 
    \item \textbf{PubChem Descriptions.} Following~\cite{MoleculeSTM, MolCA}, we employ molecular descriptions from PubChem. Due to their limited coverage (${\sim}1\%$) for molecules in reaction databases~\cite{USPTO-Applications, OpenORD}, we incorporate them exclusively for the random molecule context.
\end{itemize}

\textbf{Autoregressive Language Modeling for Interleaved Molecule-Text Sequences.} Given the input contexts above of interleaved molecules and texts, we apply language modeling loss to incrementally pretrain the LM, molecule encoder, and projector. We compute loss only for text tokens, excluding 2D molecular graph embeddings.


\begin{figure}[t]
    \centering
    \includegraphics[width=.9\linewidth]{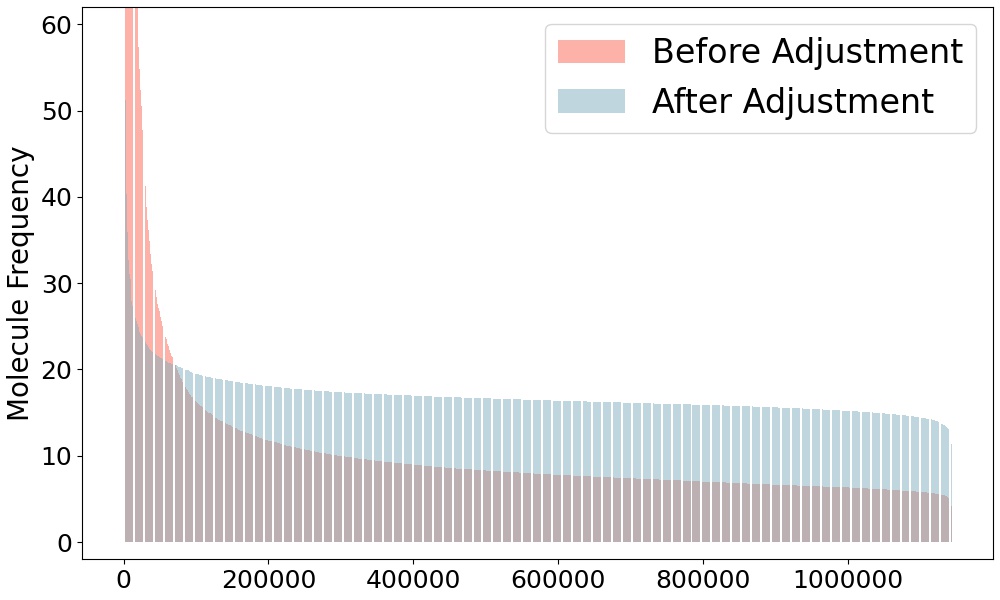}
    \caption{Distribution of molecules in the pretraining chemical reactions. For after adjustment, we conduct weighted sampling of chemical reactions matching the size of the pretraining dataset.}
    \label{fig:mol_distribution}
\end{figure}

\subsection{Balanced Sampling of Reaction Contexts}
\label{sec:balance_samp}
Figure~\ref{fig:mol_distribution} reveals a skewed distribution of molecules in chemical reactions (the red bars), with a small group of molecules appearing far more frequently than others. To address this imbalance, we develop a sampling strategy that promotes a fairer representation of molecules across reactions. This method reduces the dominance of commonly occurring molecules by adjusting 1) the sampling weight of each reaction $r$: $W(r)$, and 2) the sampling weight of each molecule $m$ within a chosen reaction $r$: $W(m|r)$, based on the equations below:

\begin{align}
    W(r) & = \frac{\sum_{m\in r}1/\text{Count}(m)}{\sum_{r'\in \mathcal{R}}\sum_{m\in r} 1/\text{Count}(m)},\label{eq:1} \\
    W(m|r) & = \frac{1/\text{Count}(m)}{\sum_{m'\in r} 1/\text{Count}(m')}, \label{eq:2}   
\end{align}
where $\mathcal{R}$ denotes the dataset of chemical reactions; $\text{Count}(m)$ denotes molecule $m$'s count in $\mathcal{R}$. 

Equation~\eqref{eq:1} sets a reaction's sampling weight inversely to the total occurrences of its molecules, favoring reactions with rare molecules; Equation~\eqref{eq:2} boosts the weights of rarer molecules within a given reaction. These weights are then applied for weighted random sampling without replacement~\cite{efraimidis2006weighted}. The blue bars in Figure~\ref{fig:mol_distribution} present the sampling frequency of molecules after adjustment, showing a flatter distribution. Implementation details are in Appendix~\ref{sec:app-exp}. 

\begin{table}[t]
    \centering
    \small
    \setlength{\tabcolsep}{1.5pt}
    \begin{tabular}{p{4.3cm}ll} \toprule
    Total reactions                                             & 2262637 & 100\%   \\ \midrule
    Too large perplexity score                                 & 329160  & 14.55\% \\
    More than one product                                      & 105577  & 4.67\%  \\
    Incomplete mapping of molecules (from chemical equation)  & 1034908 & 45.74\% \\
    Incomplete mapping of molecules (from action sequence)   & 178689  & 7.90\%  \\
    Remove duplicate reactions                                 & 254099  & 11.23\% \\
    Filter out too short actions                               & 14022   & 0.62\%  \\
    Other errors                                               & 71743   & 3.16\%  \\ \midrule
    Remaining reactions                                        & 274439  & 12.13\% \\ \bottomrule
    \end{tabular}
    \caption{Preprocessing steps and the number of samples removed at each step.}
    \label{tab:remove_count_v1}
    \end{table}
    
\begin{table}[t]
\small
\centering
\setlength{\tabcolsep}{2pt}
\begin{tabular}{lccccc} \toprule
Dataset & Total & Train & Valid & Test & Open Source \\ \midrule
\citet{smiles2actions}     & 693k  & 555k  & 69k   & 69k  & No          \\
OpenExp, Ours & 274k  & 220k  & 27k   & 27k  & Yes        \\ \bottomrule
\end{tabular}
\caption{Dataset statistics and comparison to prior work.}
\label{tab:dataset_sta}
\end{table}

\begin{table*}[t!]
    \centering
    \small
    \setlength{\tabcolsep}{2pt}
    \resizebox{\linewidth}{!}{
        \begin{tabular}{l|ccc|cccc|ccc} \toprule \textbf{Method}
            & \textbf{Validity} &             \textbf{BLEU-2} &             \textbf{BLEU-4} &             \textbf{100\%LEV} &             \textbf{90\%LEV} &             \textbf{75\%LEV} &             \textbf{50\%LEV} &             \textbf{ROUGE-1} &             \textbf{ROUGE-2} &             \textbf{ROUGE-L} \\ \midrule
            Random, among all reactions               & 63.2           & 34.5 & 19.1 & 0.0          & 0.0 & 0.0  & 13.6 & 46.6 & 18.1 & 36.4 \\
            Random, compatible pattern                & \textbf{100.0} & 37.8 & 22.1 & 0.0          & 0.0 & 0.1  & 16.5 & 47.8 & 21.0 & 38.4 \\
            Nearest neighbor                           & 76.0           & 45.0 & 30.7 & 0.6          & 6.5 & 13.0 & 38.4 & 55.7 & 29.2 & 47.0 \\
            \midrule
            TextChemT5\textsubscript{220M}           & 99.3           & 54.1 & 40.6 & 0.4          & 4.6 & 13.7 & 61.2 & 61.5 & 40.3 & 56.4 \\
            MolT5-Large\textsubscript{780M}           & 99.6           & 54.5 & 41.0 & 0.6          & 6.6 & 16.6 & 63.7 & 62.5 & 40.9 & 57.2 \\
            Galactica\textsubscript{1.3B}              & 99.9           & 53.5 & 39.5 & 0.4          & 5.7 & 13.4 & 60.5 & 60.9 & 38.6 & 55.2 \\
            MolCA, Galac\textsubscript{1.3B}                  & 99.9           & 54.9 & 41.5 & \textbf{1.0} & 9.2 & 18.9 & 65.3 & 62.5 & 40.4 & 57.0 \\
            \midrule
            ReactXT, Galac\textsubscript{1.3B}, Ours &             \textbf{100.0} &             \textbf{57.4} &             \textbf{44.0} &             \textbf{1.0} &
                \textbf{9.5} &             \textbf{22.6} &             \textbf{70.2} &             \textbf{64.4} &             \textbf{42.7} &             \textbf{58.9} \\ \bottomrule
        \end{tabular}
    }
    \vspace{-3mm}
    \caption{Comparison of experimental procedure prediction performances (\%) on the OpenExp dataset. The subscript denotes each model's parameter size. We conduct full-parameter fine-tuning for all models.}
    \vspace{-3mm}
    \label{tab:exp1_openexp}
\end{table*}

\begin{figure}[t]
    \centering
    \includegraphics[width=.9\linewidth]{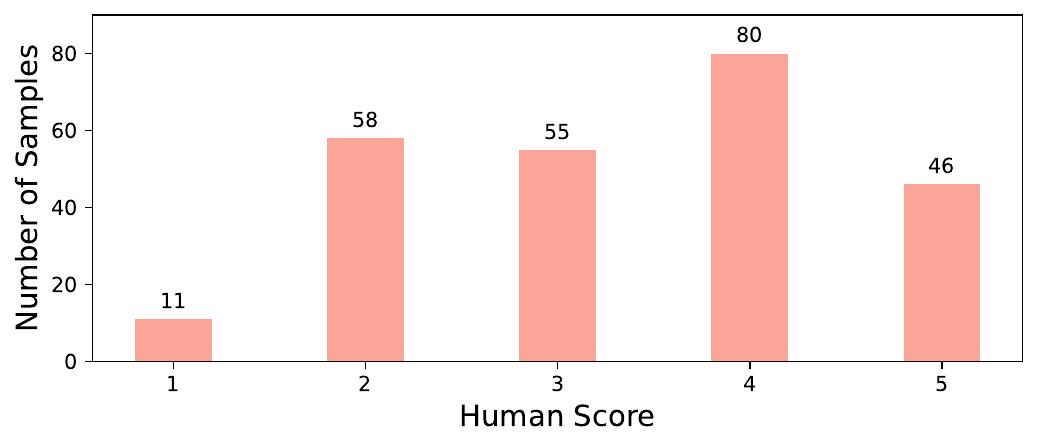}
    \vspace{-1mm}
    \caption{Human evaluations on OpenExp.}
    \vspace{-3mm}
    \label{fig:human_evaluation}
\end{figure}

\section{OpenExp: An Open-Source Dataset for Experimental Procedure Prediction}
Here we briefly introduce OpenExp's curation process and defer the details to Appendix \ref{sec:app-openexp}. OpenExp is sourced from chemical reaction databases of USPTO-Applications~\cite{USPTO-Applications} and ORD~\cite{OpenORD}. As illustrated in Figure~\ref{fig:task}, these databases include chemical reactions and the corresponding unstructured descriptions of experimental procedures. 
To convert these unstructured descriptions into structured action sequences, we first run the pragraph2action model from~\cite{TextChemT5}, and then conduct preprocessing following~\cite{smiles2actions}. The preprocessing is to remove low-quality data, eliminate duplicates, and construct molecule mapping between reactions and experimental procedures. Specific preprocessing steps are summarized in Table~\ref{tab:remove_count_v1}. An example is shown in Table \ref{tab:OpenExp_example}.

As shown in Table~\ref{tab:dataset_sta}, the final OpenExp dataset includes 274k reaction-procedure pairs. It is randomly divided into train/valid/test sets by the 8:1:1 ratio. Compared to the prior work~\cite{smiles2actions}, which is closed-source for using the commercial Pistachio database\footnote{\url{https://www.nextmovesoftware.com/pistachio}}, we open-source this dataset to assist future research. 


To obtain insights on dataset quality, we invite two graduate students in chemistry to rate the alignment between the action sequences and their original descriptions, on a scale from 1 (lowest) to 5 (highest), as depicted in Figure~\ref{fig:human_evaluation}. Briefly, of the total 250 samples evaluated, 126 ($\geq 50\%$) action sequences have at most 1 error (scores above 4), and 181 ($\geq 50\%$) action sequences have at most 2 errors (scores above 3). \shiyr{Our closer inspection shows that the one error in score-4 samples is usually a typo of material/action name, or a discrepancy of numerical value, and does not impede the overall execution. See Appendix~\ref{sec:app-human_cases} for details.}

\begin{table*}[t]
    \small
    \centering
    \begin{subtable}{\textwidth}
        \centering
        \begin{tabular}{lcccccc} \toprule
            \textbf{Method}                                       & \textbf{BLEU-2} & \textbf{BLEU-4} & \textbf{ROUGE-1} & \textbf{ROUGE-2} & \textbf{ROUGE-L} & \textbf{METEOR} \\ \midrule
            MolT5-Small\textsubscript{80M}              & 14.8   & 8.5    & 26.5    & 13.5    & 23.6    & 18.5   \\
            MolT5-Base\textsubscript{250M}              & 30.1   & 20.9   & 40.3    & 25.1    & 33.8    & 35.6   \\
            MolT5-Large\textsubscript{780M}             & 30.2   & 22.2   & 41.5    & 25.9    & 34.8    & 36.6   \\
            Galactica\textsubscript{1.3B}, LoRA ft             & 34.6   & 26.9   & 46.3    & 32.3    & 41.5    & 41.1   \\
            \midrule
            MoMu-Small\textsubscript{82M}               & 19.1   & 12.0   & 29.7    & 16.3    & 26.7    & 21.8   \\
            MoMu-Base\textsubscript{252M}               & 30.2   & 21.5   & 40.5    & 25.1    & 34.4    & 34.2   \\
            MoMu-Large\textsubscript{782M}              & 31.1   & 22.8   & 41.8    & 25.7    & 36.7    & 36.2   \\
            MolCA,   MolT5-Large\textsubscript{877M}    & 32.9   & 26.3   & 49.8    & 35.7    & 44.2    & 42.4   \\
            MolCA,   Galac\textsubscript{125M}          & 31.9   & 24.3   & 47.3    & 33.9    & 43.2    & 41.6   \\
            MolCA,   Galac\textsubscript{1.3B}, LoRA ft & 38.7   & 30.3   & 50.2    & 35.9    & 44.5    & 45.6   \\
            MolCA,   Galac\textsubscript{1.3B}, full ft*           & 39.4   & 32.2   & 52.7    & 39.4    & 47.6    & 49.2   \\ \midrule
            ReactXT,   Galac\textsubscript{1.3B}, Ours & \textbf{42.6} & \textbf{35.2} & \textbf{54.7} & \textbf{41.7} & \textbf{49.6} & \textbf{51.2} \\ \bottomrule
        \end{tabular}
        \vspace{-3pt}
        \label{tab:caption_a}
        \caption{PubChem324k dataset.}
    \end{subtable}
    \begin{subtable}{\textwidth}
        \centering
        \begin{tabular}{lcccccc} \toprule
            \textbf{Method}   & \textbf{BLEU-2} & \textbf{BLEU-4} & \textbf{ROUGE-1} & \textbf{ROUGE-2} & \textbf{ROUGE-L} & \textbf{METEOR} \\ \midrule
            MolT5-Small\textsubscript{80M}            & 51.9          & 43.6          & 62.0 & 46.9 & 56.3 & 55.1 \\
            MolT5-Base\textsubscript{250M}            & 54.0          & 45.7          & 63.4 & 48.5 & 57.8 & 56.9 \\
            MolT5-Large\textsubscript{780M}           & 59.4          & 50.8          & 65.4 & 51.0 & 59.4 & 61.4 \\
            TextChemT5\textsubscript{60M}             & 56.0          & 47.0          & 63.8 & 48.8 & 58.0 & 58.8 \\
            TextChemT5\textsubscript{220M}            & 62.5          & 54.2          & 68.2 & 54.3 & 62.2 & 64.8 \\
            \midrule
            MoMu-Small\textsubscript{82M}             & 53.2          & 44.5          & -    & -    & 56.4 & 55.7 \\
            MoMu-Base\textsubscript{252M}             & 54.9          & 46.2          & -    & -    & 57.5 & 57.6 \\
            MoMu-Large\textsubscript{782M}            & 59.9          & 51.5          & -    & -    & 59.3 & 59.7 \\
            MolCA, Galac\textsubscript{125M}          & 61.2          & 52.6          & 67.4 & 52.1 & 60.6 & 63.6 \\
            MolCA, Galac\textsubscript{1.3B}, LoRA ft & 62.0          & 53.1          & 68.1 & 53.7 & 61.8 & 65.1 \\ \midrule
            ReactXT, Galac\textsubscript{1.3B} & \textbf{62.9} & \textbf{55.0} & \textbf{69.2} & \textbf{56.0} & \textbf{63.4} & \textbf{66.4} \\ \bottomrule
        \end{tabular}
        \vspace{-3pt}
        \label{tab:caption_b}
        \caption{CheBI-20 dataset.}
    \end{subtable}
    \vspace{-7mm}
    \caption{Molecule captioning performance (\%) on the PubChem324k and CheBI-20 datasets. * denotes our re-implementation. Other baseline results are borrowed from~\cite{MolCA, TextChemT5}.}
    \vspace{-3.5mm}
    \label{tab:caption}
\end{table*}

\begin{table}[t!]
    \centering
    \small
    \setlength{\tabcolsep}{2pt}
    \resizebox{\linewidth}{!}{
        \begin{tabular}{lcccc} \toprule
            \textbf{Method}   & \textbf{Top-1} & \textbf{Top-3} & \textbf{Top-5} & \textbf{Top-10}   \\ \midrule
            MEGAN                                             & 48.1    & 70.7    & 78.4    & 86.1 \\
            AT                                                & 53.5    & -       & 81.0    & 85.7 \\
            Chemformer                                        & 54.3    & -       & 62.3    & 63.0 \\\midrule
            \multicolumn{4}{l}{\textit{Train with aug., test without aug.}} \\
            R-SMILES                                    & 51.2          & \textbf{74.9} & \textbf{81.1} & \textbf{83.0} \\
            MolT5-Large\textsubscript{780M}*            & \underline{53.9}          & 69.9          & 74.6          & 77.3          \\
            ReactXT, Galac\textsubscript{1.3B}, Ours    & \textbf{54.4} & \underline{73.6}          & \underline{78.9}          & \textbf{83.0} \\ \midrule
            \multicolumn{4}{l}{\textit{Train with aug., test with aug.}} \\
            R-SMILES                                          & \underline{56.3}          & \underline{79.2}          & \underline{86.2}          & \textbf{91.0} \\
            MolT5-Large\textsubscript{780M}*                  & 56.0          & 76.0          & 80.7          & 85.1          \\
            ReactXT, Galac\textsubscript{1.3B}, Ours          & \textbf{58.6}        & \textbf{81.1}        & \textbf{86.5}        & \textbf{91.0} \\       
            \bottomrule
        \end{tabular}
    }
    \vspace{-3mm}
    \caption{Retrosynthesis accuracies (\%) on USPTO-50K. * denotes our re-implementation. Other baselines are from~\cite{r-smiles}. In each part, \textbf{bold} denotes the best result, and \underline{underline} denotes the second best.}
    \vspace{-2mm}
    \label{tab:retrosynthesis}
\end{table}

\begin{table*}[]
\small
\centering
\setlength{\tabcolsep}{2pt}
\resizebox{\linewidth}{!}{
\begin{tabular}{l|l|cc|cc|ccc} \toprule
    \textbf{Pretrain Input Context} & \textbf{Pretrain Data Type} & \textbf{BLEU-2} & \textbf{BLEU-4} & \textbf{75\%LEV} & \textbf{50\%LEV} & \textbf{ROUGE-1} & \textbf{ROUGE-2} & \textbf{ROUGE-L} \\ \midrule
  No incremental pretrain           & -                  & 54.9 & 41.5 & 18.9 & 65.3 & 62.5 & 40.4 & 57.0 \\
  Random molecules          & reaction, sing. mol. & 56.6 & 43.2 & 20.9 & 69.4 & 63.8 & 41.9 & 58.3 \\
  Reactions w/o bal. samp.  & reaction           & 56.8 & 43.3 & 21.3 & 69.2 & 64.0 & 42.1 & 58.5 \\ 
  Reactions & reaction & 57.1 & 43.8 & 22.2 &  70.1 & 64.3 & 42.6 & \textbf{58.9} \\
  ReactXT &  reaction, sing. mol.  & \textbf{57.4} & \textbf{44.0} & \textbf{22.6} & \textbf{70.2} & \textbf{64.4} & \textbf{42.7} & \textbf{58.9} \\ \bottomrule
\end{tabular}}
\caption{Ablation study of input contexts for incrementally pretrain MolCA, Galac\textsubscript{1.3B}. Results are for experimental procedure prediction. Reactions denote both the forward reaction context and the backward reaction context.}
\label{tab:ablate}
\end{table*}

\begin{table*}[]
    \small
    \centering
    \setlength{\tabcolsep}{2pt}
    \begin{tabular}{l|cc|cc|cc|cc} \toprule
        \textbf{Backbone LM}  & \textbf{BLEU-2} & \textbf{BLEU-4} & \textbf{75\%LEV} & \textbf{50\%LEV} & \textbf{ROUGE-1} & \textbf{ROUGE-2} & \textbf{ROUGE-L} \\
        \midrule
        MolT5-Large\textsubscript{780M} & 54.5 & 41.0 & 16.6 & 63.7 & 62.5 & 40.9 & 57.2 \\
        MolT5-Large\textsubscript{780M}, ReactXT pretrain   & \textbf{55.6}          & \textbf{42.1} & \textbf{17.2} & \textbf{66.6} & \textbf{63.6} & \textbf{41.7} & \textbf{58.1} \\
        \midrule
        Galactica\textsubscript{1.3B} & 53.5 & 39.5 & 13.4 & 60.5 & 60.9 & 38.6 & 55.2 \\
        Galactica\textsubscript{1.3B}, ReactXT pretrain     & \textbf{56.5} & \textbf{43.1} & \textbf{20.8} & \textbf{68.7} & \textbf{63.7} & \textbf{41.8} & \textbf{58.2} \\
        \midrule
        MolCA, Galac\textsubscript{1.3B} & 54.9 & 41.5 & 18.9 & 65.3 & 62.5 & 40.4 & 57.0 \\
        MolCA, Galac\textsubscript{1.3B}, ReactXT pretrain  & \textbf{57.1} & \textbf{43.8} & \textbf{22.2} & \textbf{70.1} & \textbf{64.3} & \textbf{42.6} & \textbf{58.9} \\
        \bottomrule
    \end{tabular}
    \caption{Ablation study of ReactXT pretraining for experimental procedure prediction.}
    \vspace{-3mm}
    \label{tab:ablate_lm}
\end{table*}

\section{Experiment}
We empirically evaluate ReactXT across three downstream tasks, including experimental procedural prediction, molecule captioning, and retrosynthesis. Further, we include ablation studies showcasing the contributions of individual components. \shiyr{To ensure the significance of our experimental, we include statistical tests results in Appendix ~\ref{sec:app-t_test}.}

\subsection{Experimental Setting}
ReactXT is initialized by the stage-2 checkpoint of MolCA\textsubscript{1.3B}~\cite{MolCA}, if not specially noted. It is then pretrained using our proposed method, and subsequently finetuned for each downstream dataset separately. The context length $k$ is $4$. We employ full-parameter tuning for pretraining and finetuning. More details are in Appendix~\ref{sec:app-exp}.


\textbf{ReactXT's Pretraining Dataset.} Our pretrain dataset includes PubChem324k's pretrain subset~\cite{MolCA}, which includes 298k molecule-text pairs, and 1.11 million chemical reactions from the USPTO-Applications~\cite{USPTO-Applications} and ORD~\cite{OpenORD} databases. For molecules in reactions, we obtain their patent abstracts and molecular properties following Section~\ref{sec:input_context}. To prevent information leakage, we have excluded 54k reactions that appear in the valid/test sets of the downstream datasets (\ie OpenExp, USPTO-50K~\cite{uspto-50k}) from the initial collection of 1.16 million reactions. See Appendix \ref{sec:app-pretrain_dataset} for more details.

\textbf{Baselines.} We compare ReactXT with the state-of-the-art LMs in science domain, including Galactica \cite{Galactica}, MolT5~\cite{MolT5}, TextChemT5~\cite{TextChemT5}, and MolCA~\cite{MolCA}. For retrosynthesis and forward reaction prediction tasks, we also compare with task-specific LMs: R-SMILES~\cite{r-smiles}, AT~\cite{AT}, MEGAN~\cite{megan}, and Chemformer~\cite{chemformer}. For captioning, we additionally compare against MoMu~\cite{MoMu}. 

\subsection{Experimental Procedure Prediction}
\vspace{-2mm}
Following~\cite{smiles2actions}, we employ the following evaluation metrics: Validity, which checks the syntactical correctness of the action sequence; machine-translation metrics BLUE~\cite{BLEU} and ROUGE~\cite{ROUGE}; and the normalized Levenshtein similarity~\cite{levenshtein1966binary}. Specifically, 90\%LEV denotes the proportion of predictions with a normalized Levenshtein score larger than 0.9. The three naive baselines based on random sampling and nearest neighbor are borrowed from~\cite{smiles2actions}. See Appendix~\ref{sec:app-exp} for details.

Table~\ref{tab:exp1_openexp} presents the performances. We can observe that ReactXT consistently outperforms baselines across all metrics. Specifically, it surpasses baselines by 2.2\% for BLEU-2 and 3.3\% for 75\%LEV, demonstrating ReactXT's effectiveness for text-based reaction understanding. 

\subsection{Molecule Captioning}
To evaluate ReactXT's ability to understand single-molecules, we present its performances of molecule captioning on the PubChem324k~\cite{MolCA} and CheBI-20~\cite{MolT5} datasets. We report metrics of BLEU~\cite{BLEU}, ROUGE~\cite{ROUGE}, and METEOR~\cite{METEOR}. 

Table~\ref{tab:caption} presents the captioning performances. We can observe that ReactXT consistently outperforms the baselines. Specifically, ReactXT shows improvements of 3.2\% BLEU-2 and 2.3\% ROUGE-2 scores on PubChem324k, and 1.7\% ROUGE-2 on CheBI-20. These improvements underscore the effectiveness of our pretraining method for enhancing understanding of individual molecules.

\subsection{Retrosynthesis}
Retrosynthesis is to predict the reactant molecules given the product molecules. For this task, we employ the evaluation metrics of top-k accuracy, which measures the percentage of exact match to the ground truth in the top-k predictions. Following~\cite{r-smiles}, we conduct self-supervised pretraining on the USPTO-full\cite{GLN} dataset and use the root-aligned augmentations of SMILES during training and testing. Additionally, we report performances of testing without these augmentations.

\shiyr{Table~\ref{tab:retrosynthesis} presents the results. ReactXT outperforms R-SMILES across all metrics when testing with augmentations. Notably, the improvement in top-1 accuracy is particularly significant, achieving a 2.3\% increase over the second best value. Regardless of whether test set data augmentation is applied, ReactXT achieves better top-k accuracies than MolT5-Large, which is also a multimodal LM. These performance improvements stem from ReactXT's use of reactions for pretraining, rather than individual molecules.}

\subsection{Ablation Study}
In this section, we conduct ablation studies to show the impact of different pretrain data types and backbone LMs in our method.

\textbf{Pretrain Data Type.} We ablate the key components of ReactXT, using the baseline of MolCA, Galac\textsubscript{1.3B} without incremental pretraining. Table~\ref{tab:ablate} presents the results. Specifically, we compare three variants of ReactXT: 1) pretraining with solely the random molecule contexts using the same pretrain dataset; 2) pretraining with forward and backward reaction contexts without the random molecule context; and 3) applying uniform sampling on reaction contexts instead of balanced sampling.

We can observe that 1) ReactXT's full model shows the best performance, showing its performance is the integrated contribution of all components; 2) applying random molecule contexts alone improves upon the baseline, underscoring the valuable textual knowledge from our meticulously crafted pretraining dataset; 3) incorporating reaction contexts yields better results than random molecule contexts, highlighting the benefits of learning reaction knowledge during pretraining; and 4) balanced sampling improves the performance upon uniform sampling. 

\textbf{Backbone LMs.}
We conduct ablation studies on the backbone LMs. This study involves three different molecular-text LMs: 1) MolCA, which represents molecules using both 1D SMILES and 2D graphs, based on a decoder-only architecture; 2) Galactica, which represents molecules using 1D SMILES, based on a decoder-only architecture; and 3) MolT5, which represents molecules using 1D SMILES, based on an encoder-decoder architecture. The experimental results are presented in Table~\ref{tab:ablate_lm}. We can observe that the ReactXT pretraining scheme achieves consistent performance improvements, regardless of the backbone language model used.
\section{Conclusion and Future Works}
In this work, we explore reaction-text modeling to empower reaction-relevant tasks with textual interfaces and knowledge. We present ReactXT, a pretraining method to learn chemical reactions within the context of the corresponding molecular textual descriptions. Additionally, we propose a new dataset OpenExp to support open-source research for experimental procedure prediction. ReactXT establishes the best performances across tasks of experimental procedure prediction and molecule captioning. It presents competitive performances for retrosynthesis.

In future work, we plan to apply LMs to learn the interactions among large molecules (\eg proteins and nucleic acids), or introduce molecules' dynamics and 3D spatial structures for better molecule-language understanding~\cite{TEDMol}. We are also interested in exploring molecular LMs for OOD generalization~\cite{fangjf2023eva,fangjf2023exgc}.

\section*{Limitations}
In this and also the previous work~\cite{smiles2actions}, the evaluation for experimental procedure prediction is constrained to the comparison between the predictions and the reference action sequences. While improving this metric does reflect the improvement in experimental design, it should be acknowledged that the evaluation of real-world chemical experiments is preferred for the developed models in future. For this purpose, the methods on automated chemistry pipelines~\cite{auto_research, coley2019robotic, nicolaou2020context} can be potentially considered.

Another limitation or future direction is improving the action space defined in our proposed OpenExp dataset, aiming to cover a wider range of chemical experiments. For example, the action of `Purify' is absent; and the action of `Concentration' can be refined into operations such as `Evaporation' and `Pressurize' for clearer instructions of chemical experiments. 


\section*{Potential Ethics Impact}
In this study, the proposed method and dataset focus on chemical reactions and molecules, and include no human subjects. Consequently, we believe this study presents no direct ethical concerns. However, the inclusion of LMs in our study does raise potential issues, as LMs can be misused to produce incorrect or biased information. Therefore, the ethical implications of our work align with those common to LM research, emphasizing the need for responsible use and application of LMs.

\section*{Acknowledgement}
This research is supported by the National Science and Technology Major Project (2023ZD0121102), National Natural Science Foundation of China (92270114). This research is partially supported by the National Research Foundation Singapore under the AI Singapore Programme (AISG Award No: AISG2-TC-2023-010-SGIL), the Singapore Ministry of Education Academic Research Fund Tier 1 (Award No: T1 251RES2207) and the Google Cloud Research Credits program with the award (Q4MJ-YH1K-3MVX-FP6Q). This research is supported by NExT Research Center.

\bibliography{reference}

\begin{thebibliography}{60}
\expandafter\ifx\csname natexlab\endcsname\relax\def\natexlab#1{#1}\fi

\bibitem[{Atkins and Jones(2007)}]{atkins2007chemical}
Peter Atkins and Loretta Jones. 2007.
\newblock \emph{Chemical principles: The quest for insight}.
\newblock Macmillan.

\bibitem[{Banerjee and Lavie(2005)}]{METEOR}
Satanjeev Banerjee and Alon Lavie. 2005.
\newblock {METEOR:} an automatic metric for {MT} evaluation with improved correlation with human judgments.
\newblock In \emph{IEEvaluation@ACL}, pages 65--72. Association for Computational Linguistics.

\bibitem[{Berglund et~al.(2023)Berglund, Tong, Kaufmann, Balesni, Stickland, Korbak, and Evans}]{ReversalCurse}
Lukas Berglund, Meg Tong, Max Kaufmann, Mikita Balesni, Asa~Cooper Stickland, Tomasz Korbak, and Owain Evans. 2023.
\newblock The reversal curse: Llms trained on" a is b" fail to learn" b is a".
\newblock \emph{arXiv preprint arXiv:2309.12288}.

\bibitem[{Boiko et~al.(2023)Boiko, MacKnight, Kline, and Gomes}]{auto_research}
Daniil~A Boiko, Robert MacKnight, Ben Kline, and Gabe Gomes. 2023.
\newblock Autonomous chemical research with large language models.
\newblock \emph{Nature}, 624(7992):570--578.

\bibitem[{Born and Manica(2023)}]{RegressionTransformer}
Jannis Born and Matteo Manica. 2023.
\newblock Regression transformer enables concurrent sequence regression and generation for molecular language modelling.
\newblock \emph{Nat. Mac. Intell.}, 5(4):432--444.

\bibitem[{Bran et~al.(2023)Bran, Cox, Schilter, Baldassari, White, and Schwaller}]{chemcrow}
Andres~M Bran, Sam Cox, Oliver Schilter, Carlo Baldassari, Andrew White, and Philippe Schwaller. 2023.
\newblock Augmenting large language models with chemistry tools.
\newblock In \emph{NeurIPS 2023 AI for Science Workshop}.

\bibitem[{Burger et~al.(2020)Burger, Maffettone, Gusev, Aitchison, Bai, Wang, Li, Alston, Li, Clowes et~al.}]{MobileChemist}
Benjamin Burger, Phillip~M Maffettone, Vladimir~V Gusev, Catherine~M Aitchison, Yang Bai, Xiaoyan Wang, Xiaobo Li, Ben~M Alston, Buyi Li, Rob Clowes, et~al. 2020.
\newblock A mobile robotic chemist.
\newblock \emph{Nature}, 583(7815):237--241.

\bibitem[{Chen and Jung(2021)}]{localretro}
Shuan Chen and Yousung Jung. 2021.
\newblock Deep retrosynthetic reaction prediction using local reactivity and global attention.
\newblock \emph{JACS Au}, 1(10):1612--1620.

\bibitem[{Chithrananda et~al.(2020)Chithrananda, Grand, and Ramsundar}]{chemberta}
Seyone Chithrananda, Gabriel Grand, and Bharath Ramsundar. 2020.
\newblock Chemberta: large-scale self-supervised pretraining for molecular property prediction.
\newblock \emph{arXiv preprint arXiv:2010.09885}.

\bibitem[{Christofidellis et~al.(2023)Christofidellis, Giannone, Born, Winther, Laino, and Manica}]{TextChemT5}
Dimitrios Christofidellis, Giorgio Giannone, Jannis Born, Ole Winther, Teodoro Laino, and Matteo Manica. 2023.
\newblock Unifying molecular and textual representations via multi-task language modelling.
\newblock In \emph{{ICML}}.

\bibitem[{Coley et~al.(2019)Coley, Thomas~III, Lummiss, Jaworski, Breen, Schultz, Hart, Fishman, Rogers, Gao et~al.}]{coley2019robotic}
Connor~W Coley, Dale~A Thomas~III, Justin~AM Lummiss, Jonathan~N Jaworski, Christopher~P Breen, Victor Schultz, Travis Hart, Joshua~S Fishman, Luke Rogers, Hanyu Gao, et~al. 2019.
\newblock A robotic platform for flow synthesis of organic compounds informed by ai planning.
\newblock \emph{Science}, 365(6453):eaax1566.

\bibitem[{Dai et~al.(2019)Dai, Li, Coley, Dai, and Song}]{GLN}
Hanjun Dai, Chengtao Li, Connor Coley, Bo~Dai, and Le~Song. 2019.
\newblock Retrosynthesis prediction with conditional graph logic network.
\newblock \emph{Advances in Neural Information Processing Systems}, 32.

\bibitem[{Edwards et~al.(2022)Edwards, Lai, Ros, Honke, Cho, and Ji}]{MolT5}
Carl Edwards, Tuan~Manh Lai, Kevin Ros, Garrett Honke, Kyunghyun Cho, and Heng Ji. 2022.
\newblock Translation between molecules and natural language.
\newblock In \emph{{EMNLP}}, pages 375--413. Association for Computational Linguistics.

\bibitem[{Edwards et~al.(2021)Edwards, Zhai, and Ji}]{Text2Mol}
Carl Edwards, ChengXiang Zhai, and Heng Ji. 2021.
\newblock Text2mol: Cross-modal molecule retrieval with natural language queries.
\newblock In \emph{{EMNLP} {(1)}}, pages 595--607. Association for Computational Linguistics.

\bibitem[{Efraimidis and Spirakis(2006)}]{efraimidis2006weighted}
Pavlos~S Efraimidis and Paul~G Spirakis. 2006.
\newblock Weighted random sampling with a reservoir.
\newblock \emph{Information processing letters}, 97(5):181--185.

\bibitem[{Fang et~al.(2024{\natexlab{a}})Fang, Li, Sui, Gao, Zhang, Wang, Wang, and He}]{fangjf2023exgc}
Junfeng Fang, Xinglin Li, Yongduo Sui, Yuan Gao, Guibin Zhang, Kun Wang, Xiang Wang, and Xiangnan He. 2024{\natexlab{a}}.
\newblock Exgc: Bridging efficiency and explainability in graph condensation.
\newblock In \emph{{WWW}}. {ACM}.

\bibitem[{Fang et~al.(2023{\natexlab{a}})Fang, Liu, Gao, Liu, Zhang, Wang, and He}]{fangjf2023eva}
Junfeng Fang, Wei Liu, Yuan Gao, Zemin Liu, An~Zhang, Xiang Wang, and Xiangnan He. 2023{\natexlab{a}}.
\newblock Evaluating post-hoc explanations for graph neural networks via robustness analysis.
\newblock In \emph{Thirty-seventh Conference on Neural Information Processing Systems}.

\bibitem[{Fang et~al.(2024{\natexlab{b}})Fang, Zhang, Wu, Yang, Liu, Li, Wang, Du, and Wang}]{moltc}
Junfeng Fang, Shuai Zhang, Chang Wu, Zhengyi Yang, Zhiyuan Liu, Sihang Li, Kun Wang, Wenjie Du, and Xiang Wang. 2024{\natexlab{b}}.
\newblock {M}ol{TC}: Towards molecular relational modeling in language models.
\newblock \emph{arXiv preprint arXiv:2402.03781}.

\bibitem[{Fang et~al.(2023{\natexlab{b}})Fang, Liang, Zhang, Liu, Huang, Chen, Fan, and Chen}]{MolInstructions}
Yin Fang, Xiaozhuan Liang, Ningyu Zhang, Kangwei Liu, Rui Huang, Zhuo Chen, Xiaohui Fan, and Huajun Chen. 2023{\natexlab{b}}.
\newblock Mol-instructions: {A} large-scale biomolecular instruction dataset for large language models.
\newblock \emph{CoRR}, abs/2306.08018.

\bibitem[{Gao et~al.(2018)Gao, Struble, Coley, Wang, Green, and Jensen}]{gao2018using}
Hanyu Gao, Thomas~J Struble, Connor~W Coley, Yuran Wang, William~H Green, and Klavs~F Jensen. 2018.
\newblock Using machine learning to predict suitable conditions for organic reactions.
\newblock \emph{ACS central science}, 4(11):1465--1476.

\bibitem[{Guo et~al.(2023)Guo, Guo, Nan, Liang, Guo, Chawla, Wiest, and Zhang}]{GPTChemBench}
Taicheng Guo, Kehan Guo, Bozhao Nan, Zhengwen Liang, Zhichun Guo, Nitesh~V. Chawla, Olaf Wiest, and Xiangliang Zhang. 2023.
\newblock What indeed can {GPT} models do in chemistry? {A} comprehensive benchmark on eight tasks.
\newblock \emph{CoRR}, abs/2305.18365.

\bibitem[{Hu et~al.(2020)Hu, Liu, Gomes, Zitnik, Liang, Pande, and Leskovec}]{pretrain_gnn}
Weihua Hu, Bowen Liu, Joseph Gomes, Marinka Zitnik, Percy Liang, Vijay Pande, and Jure Leskovec. 2020.
\newblock Strategies for pre-training graph neural networks.
\newblock In \emph{ICLR}.

\bibitem[{Irwin et~al.(2022)Irwin, Dimitriadis, He, and Bjerrum}]{chemformer}
Ross Irwin, Spyridon Dimitriadis, Jiazhen He, and Esben~Jannik Bjerrum. 2022.
\newblock Chemformer: a pre-trained transformer for computational chemistry.
\newblock \emph{Machine Learning: Science and Technology}, 3(1):015022.

\bibitem[{Kearnes et~al.(2021)Kearnes, Maser, Wleklinski, Kast, Doyle, Dreher, Hawkins, Jensen, and Coley}]{OpenORD}
Steven~M Kearnes, Michael~R Maser, Michael Wleklinski, Anton Kast, Abigail~G Doyle, Spencer~D Dreher, Joel~M Hawkins, Klavs~F Jensen, and Connor~W Coley. 2021.
\newblock The open reaction database.
\newblock \emph{Journal of the American Chemical Society}, 143(45):18820--18826.

\bibitem[{Krenn et~al.(2020)Krenn, H{\"{a}}se, Nigam, Friederich, and Aspuru{-}Guzik}]{selfies}
Mario Krenn, Florian H{\"{a}}se, AkshatKumar Nigam, Pascal Friederich, and Al{\'{a}}n Aspuru{-}Guzik. 2020.
\newblock \href {https://doi.org/10.1088/2632-2153/ABA947} {Self-referencing embedded strings {(SELFIES):} {A} 100{\%} robust molecular string representation}.
\newblock \emph{Mach. Learn. Sci. Technol.}, 1(4):45024.

\bibitem[{Levenshtein et~al.(1966)}]{levenshtein1966binary}
Vladimir~I Levenshtein et~al. 1966.
\newblock Binary codes capable of correcting deletions, insertions, and reversals.
\newblock In \emph{Soviet physics doklady}, volume~10, pages 707--710. Soviet Union.

\bibitem[{Li et~al.(2023)Li, Li, Savarese, and Hoi}]{BLIP2}
Junnan Li, Dongxu Li, Silvio Savarese, and Steven C.~H. Hoi. 2023.
\newblock {BLIP-2:} bootstrapping language-image pre-training with frozen image encoders and large language models.
\newblock \emph{CoRR}, abs/2301.12597.

\bibitem[{Li et~al.(2024)Li, Liu, Luo, Wang, He, Kawaguchi, Chua, and Tian}]{3dmolm}
Sihang Li, Zhiyuan Liu, Yanchen Luo, Xiang Wang, Xiangnan He, Kenji Kawaguchi, Tat-Seng Chua, and Qi~Tian. 2024.
\newblock \href {https://openreview.net/forum?id=xI4yNlkaqh} {3d-molm: Towards 3d molecule-text interpretation in language models}.
\newblock In \emph{{ICLR}}.

\bibitem[{Lin(2004)}]{ROUGE}
Chin-Yew Lin. 2004.
\newblock Rouge: A package for automatic evaluation of summaries.
\newblock In \emph{Text summarization branches out}, pages 74--81.

\bibitem[{Liu et~al.(2017)Liu, Ramsundar, Kawthekar, Shi, Gomes, Luu~Nguyen, Ho, Sloane, Wender, and Pande}]{seq2seq}
Bowen Liu, Bharath Ramsundar, Prasad Kawthekar, Jade Shi, Joseph Gomes, Quang Luu~Nguyen, Stephen Ho, Jack Sloane, Paul Wender, and Vijay Pande. 2017.
\newblock Retrosynthetic reaction prediction using neural sequence-to-sequence models.
\newblock \emph{ACS central science}, 3(10):1103--1113.

\bibitem[{Liu et~al.(2023{\natexlab{a}})Liu, Li, Wu, and Lee}]{llava}
Haotian Liu, Chunyuan Li, Qingyang Wu, and Yong~Jae Lee. 2023{\natexlab{a}}.
\newblock Visual instruction tuning.
\newblock \emph{arXiv preprint arXiv:2304.08485}.

\bibitem[{Liu et~al.(2022)Liu, Nie, Wang, Lu, Qiao, Liu, Tang, Xiao, and Anandkumar}]{MoleculeSTM}
Shengchao Liu, Weili Nie, Chengpeng Wang, Jiarui Lu, Zhuoran Qiao, Ling Liu, Jian Tang, Chaowei Xiao, and Anima Anandkumar. 2022.
\newblock Multi-modal molecule structure-text model for text-based retrieval and editing.
\newblock \emph{CoRR}, abs/2212.10789.

\bibitem[{Liu et~al.(2023{\natexlab{b}})Liu, Li, Luo, Fei, Cao, Kawaguchi, Wang, and Chua}]{MolCA}
Zhiyuan Liu, Sihang Li, Yanchen Luo, Hao Fei, Yixin Cao, Kenji Kawaguchi, Xiang Wang, and Tat{-}Seng Chua. 2023{\natexlab{b}}.
\newblock {M}ol{CA}: Molecular graph-language modeling with cross-modal projector and uni-modal adapter.
\newblock In \emph{{EMNLP}}, pages 15623--15638. Association for Computational Linguistics.

\bibitem[{Liu et~al.(2023{\natexlab{c}})Liu, Shi, Zhang, Zhang, Kawaguchi, Wang, and Chua}]{simsgt}
Zhiyuan Liu, Yaorui Shi, An~Zhang, Enzhi Zhang, Kenji Kawaguchi, Xiang Wang, and Tat-Seng Chua. 2023{\natexlab{c}}.
\newblock \href {https://openreview.net/forum?id=fWLf8DV0fI} {Rethinking tokenizer and decoder in masked graph modeling for molecules}.
\newblock In \emph{{NeurIPS}}.

\bibitem[{Lowe(2017)}]{USPTO-Applications}
Daniel Lowe. 2017.
\newblock \href {https://doi.org/10.6084/m9.figshare.5104873.v1} {{Chemical reactions from US patents (1976-Sep2016)}}.

\bibitem[{Luo et~al.(2023)Luo, Li, Liu, Wu, Yang, He, Wang, and Tian}]{TEDMol}
Yanchen Luo, Sihang Li, Zhiyuan Liu, Jiancan Wu, Zhengyi Yang, Xiangnan He, Xiang Wang, and Qi~Tian. 2023.
\newblock Text-guided diffusion model for 3d molecule generation.

\bibitem[{Nicolaou et~al.(2020)Nicolaou, Watson, Lemasters, Masquelin, and Wang}]{nicolaou2020context}
Christos~A. Nicolaou, Ian~A. Watson, Mark Lemasters, Thierry Masquelin, and Ji{-}Bo Wang. 2020.
\newblock Context aware data-driven retrosynthetic analysis.
\newblock \emph{J. Chem. Inf. Model.}, 60(6):2728--2738.

\bibitem[{OpenAI(2023)}]{GPT4}
OpenAI. 2023.
\newblock {GPT-4} technical report.
\newblock \emph{CoRR}, abs/2303.08774.

\bibitem[{Papineni et~al.(2002)Papineni, Roukos, Ward, and Zhu}]{BLEU}
Kishore Papineni, Salim Roukos, Todd Ward, and Wei{-}Jing Zhu. 2002.
\newblock Bleu: a method for automatic evaluation of machine translation.
\newblock In \emph{{ACL}}, pages 311--318. {ACL}.

\bibitem[{Radford et~al.(2021)Radford, Kim, Hallacy, Ramesh, Goh, Agarwal, Sastry, Askell, Mishkin, Clark, Krueger, and Sutskever}]{CLIP}
Alec Radford, Jong~Wook Kim, Chris Hallacy, Aditya Ramesh, Gabriel Goh, Sandhini Agarwal, Girish Sastry, Amanda Askell, Pamela Mishkin, Jack Clark, Gretchen Krueger, and Ilya Sutskever. 2021.
\newblock Learning transferable visual models from natural language supervision.
\newblock In \emph{{ICML}}, volume 139 of \emph{Proceedings of Machine Learning Research}, pages 8748--8763. {PMLR}.

\bibitem[{Rajan et~al.(2021)Rajan, Zielesny, and Steinbeck}]{stout}
Kohulan Rajan, Achim Zielesny, and Christoph Steinbeck. 2021.
\newblock Stout: Smiles to iupac names using neural machine translation.
\newblock \emph{Journal of Cheminformatics}, 13(1):1--14.

\bibitem[{Sacha et~al.(2021)Sacha, B{\l}az, Byrski, Dabrowski-Tumanski, Chrominski, Loska, W{\l}odarczyk-Pruszynski, and Jastrzebski}]{megan}
Miko{\l}aj Sacha, Miko{\l}aj B{\l}az, Piotr Byrski, Pawe{\l} Dabrowski-Tumanski, Miko{\l}aj Chrominski, Rafa{\l} Loska, Pawe{\l} W{\l}odarczyk-Pruszynski, and Stanis{\l}aw Jastrzebski. 2021.
\newblock Molecule edit graph attention network: modeling chemical reactions as sequences of graph edits.
\newblock \emph{Journal of Chemical Information and Modeling}, 61(7):3273--3284.

\bibitem[{Schneider et~al.(2016)Schneider, Stiefl, and Landrum}]{uspto-50k}
Nadine Schneider, Nikolaus Stiefl, and Gregory~A Landrum. 2016.
\newblock What’s what: The (nearly) definitive guide to reaction role assignment.
\newblock \emph{Journal of chemical information and modeling}, 56(12):2336--2346.

\bibitem[{Schwaller et~al.(2019)Schwaller, Probst, Vaucher, Nair, Laino, and Reymond}]{schwaller2019data}
Philippe Schwaller, Daniel Probst, Alain~C Vaucher, Vishnu~H Nair, Teodoro Laino, and Jean-Louis Reymond. 2019.
\newblock Data-driven chemical reaction classification, fingerprinting and clustering using attention-based neural networks.
\newblock \emph{ChemRxiv}.

\bibitem[{Schwaller et~al.(2022)Schwaller, Vaucher, Laplaza, Bunne, Krause, Corminboeuf, and Laino}]{schwaller2022machine}
Philippe Schwaller, Alain~C Vaucher, Ruben Laplaza, Charlotte Bunne, Andreas Krause, Clemence Corminboeuf, and Teodoro Laino. 2022.
\newblock Machine intelligence for chemical reaction space.
\newblock \emph{Wiley Interdisciplinary Reviews: Computational Molecular Science}, 12(5):e1604.

\bibitem[{Shi et~al.(2020)Shi, Xu, Guo, Zhang, and Tang}]{g2g}
Chence Shi, Minkai Xu, Hongyu Guo, Ming Zhang, and Jian Tang. 2020.
\newblock A graph to graphs framework for retrosynthesis prediction.
\newblock In \emph{Proceedings of the 37th International Conference on Machine Learning, {ICML} 2020, 13-18 July 2020, Virtual Event}, volume 119 of \emph{Proceedings of Machine Learning Research}, pages 8818--8827. {PMLR}.

\bibitem[{Shi et~al.(2023)Shi, Zhang, Zhang, Liu, and Wang}]{relm}
Yaorui Shi, An~Zhang, Enzhi Zhang, Zhiyuan Liu, and Xiang Wang. 2023.
\newblock \href {https://aclanthology.org/2023.findings-emnlp.366} {{R}e{LM}: Leveraging language models for enhanced chemical reaction prediction}.
\newblock In \emph{Findings of the Association for Computational Linguistics: {EMNLP} 2023, Singapore, December 6-10, 2023}, pages 5506--5520. Association for Computational Linguistics.

\bibitem[{Su et~al.(2022)Su, Du, Yang, Zhou, Li, Rao, Sun, Lu, and Wen}]{MoMu}
Bing Su, Dazhao Du, Zhao Yang, Yujie Zhou, Jiangmeng Li, Anyi Rao, Hao Sun, Zhiwu Lu, and Ji{-}Rong Wen. 2022.
\newblock A molecular multimodal foundation model associating molecule graphs with natural language.
\newblock \emph{CoRR}, abs/2209.05481.

\bibitem[{Taylor et~al.(2022)Taylor, Kardas, Cucurull, Scialom, Hartshorn, Saravia, Poulton, Kerkez, and Stojnic}]{Galactica}
Ross Taylor, Marcin Kardas, Guillem Cucurull, Thomas Scialom, Anthony Hartshorn, Elvis Saravia, Andrew Poulton, Viktor Kerkez, and Robert Stojnic. 2022.
\newblock Galactica: {A} large language model for science.
\newblock \emph{CoRR}, abs/2211.09085.

\bibitem[{Tetko et~al.(2020)Tetko, Karpov, Van~Deursen, and Godin}]{AT}
Igor~V Tetko, Pavel Karpov, Ruud Van~Deursen, and Guillaume Godin. 2020.
\newblock State-of-the-art augmented nlp transformer models for direct and single-step retrosynthesis.
\newblock \emph{Nature communications}, 11(1):5575.

\bibitem[{Ucak et~al.(2022)Ucak, Ashyrmamatov, Ko, and Lee}]{RetroTRAE}
Umit~V Ucak, Islambek Ashyrmamatov, Junsu Ko, and Juyong Lee. 2022.
\newblock Retrosynthetic reaction pathway prediction through neural machine translation of atomic environments.
\newblock \emph{Nature communications}, 13(1):1186.

\bibitem[{Vaucher et~al.(2021)Vaucher, Schwaller, Geluykens, Nair, Iuliano, and Laino}]{smiles2actions}
Alain~C Vaucher, Philippe Schwaller, Joppe Geluykens, Vishnu~H Nair, Anna Iuliano, and Teodoro Laino. 2021.
\newblock Inferring experimental procedures from text-based representations of chemical reactions.
\newblock \emph{Nature communications}, 12(1):2573.

\bibitem[{Vaucher et~al.(2020)Vaucher, Zipoli, Geluykens, Nair, Schwaller, and Laino}]{paragraph2actions}
Alain~C Vaucher, Federico Zipoli, Joppe Geluykens, Vishnu~H Nair, Philippe Schwaller, and Teodoro Laino. 2020.
\newblock Automated extraction of chemical synthesis actions from experimental procedures.
\newblock \emph{Nature communications}, 11(1):3601.

\bibitem[{Weininger(1988)}]{SMILES}
David Weininger. 1988.
\newblock Smiles, a chemical language and information system. 1. introduction to methodology and encoding rules.
\newblock \emph{J. Chem. Inf. Comput. Sci.}, 28(1):31--36.

\bibitem[{Wells(2012)}]{wells2012structural}
Alexander~Frank Wells. 2012.
\newblock \emph{Structural inorganic chemistry}.
\newblock Oxford university press.

\bibitem[{Yan et~al.(2020)Yan, Ding, Zhao, Zheng, Yang, Yu, and Huang}]{RetroXpert}
Chaochao Yan, Qianggang Ding, Peilin Zhao, Shuangjia Zheng, Jinyu Yang, Yang Yu, and Junzhou Huang. 2020.
\newblock Retroxpert: Decompose retrosynthesis prediction like {A} chemist.
\newblock In \emph{NeurIPS}.

\bibitem[{You et~al.(2020)You, Chen, Sui, Chen, Wang, and Shen}]{GraphCL}
Yuning You, Tianlong Chen, Yongduo Sui, Ting Chen, Zhangyang Wang, and Yang Shen. 2020.
\newblock Graph contrastive learning with augmentations.
\newblock In \emph{NeurIPS}.

\bibitem[{Zeng et~al.(2023)Zeng, Nie, Ding, Ding, Ye, Yang, Sun, Weinan, Zhu, and Liu}]{description2instructions}
Zheni Zeng, Yi-Chen Nie, Ning Ding, Qian-Jun Ding, Wei-Ting Ye, Cheng Yang, Maosong Sun, E~Weinan, Rong Zhu, and Zhiyuan Liu. 2023.
\newblock Transcription between human-readable synthetic descriptions and machine-executable instructions: an application of the latest pre-training technology.
\newblock \emph{Chemical Science}, 14(35):9360--9373.

\bibitem[{Zeng et~al.(2022)Zeng, Yao, Liu, and Sun}]{KVPLM}
Zheni Zeng, Yuan Yao, Zhiyuan Liu, and Maosong Sun. 2022.
\newblock A deep-learning system bridging molecule structure and biomedical text with comprehension comparable to human professionals.
\newblock \emph{Nature communications}, 13(1):862.

\bibitem[{Zhong et~al.(2022)Zhong, Song, Feng, Liu, Jia, Yao, Wu, Hou, and Song}]{r-smiles}
Zipeng Zhong, Jie Song, Zunlei Feng, Tiantao Liu, Lingxiang Jia, Shaolun Yao, Min Wu, Tingjun Hou, and Mingli Song. 2022.
\newblock Root-aligned smiles: a tight representation for chemical reaction prediction.
\newblock \emph{Chemical Science}, 13(31):9023--9034.

\end{thebibliography}

\clearpage
\appendix
\section{Dataset Details}
\label{sec:app-dataset}

\subsection{Collection and Preprocessing of OpenExp}
\label{sec:app-openexp}
OpenExp is compiled from the raw data from the two following sources:
\begin{itemize}[leftmargin=*]
\item \textbf{USPTO-Applications}~\cite{USPTO-Applications}. This dataset comprises records of 1.94 million reactions and their corresponding applications from the United States Patent and Trademark Office (USPTO) published between 2001 and September 2016. We download the raw XML files from the Figshare website \footnote{\url{https://figshare.com/articles/dataset/Chemical_reactions_from_US_patents_1976-Sep2016_/5104873?file=8664370}}. For each reaction in this dataset, we extract its key information from four elements: \verb|<productList>|, which contains the products of the reaction; \verb|<reactantList>|, detailing the reactants; \verb|<spectatorList>|, encompassing the catalysts and solvents; and \verb|<dl:paragraphText>|, which provides a textual description of the experimental procedures.

\item \textbf{Open Reaction Database}~\cite{OpenORD}. The ORD \footnote{\url{https://open-reaction-database.org}} dataset contains over 2 million chemical reactions, which include detailed records of reaction conditions and experimental procedures. It includes data from the USPTO applications (2001-2016 Sep), USPTO-granted patents (1976-2016 Sep), and experimental records from chemical literature.
\end{itemize}

\textbf{Paragraph2Action.} As illustrated in Figure~\ref{fig:task}, these databases include chemical reactions and the corresponding unstructured descriptions of experimental procedures. The unstructured nature of these descriptions poses a significant challenge to 1) automate chemical synthesis with robots~\cite{paragraph2actions, MobileChemist}; and 2) apply ML methods to predict experimental procedures of unseen reactions. To address this, the task of paragraph2action~\cite{paragraph2actions, description2instructions} is proposed, aiming to convert unstructured experimental procedure descriptions into structured, step-by-step instructions with pre-defined actions. In this study, we leverage the action space defined by~\cite{paragraph2actions,smiles2actions}, and the pragraph2action model released by~\cite{TextChemT5}.


\begin{table}[t]
    \centering
    \small
    \setlength{\tabcolsep}{1.5pt}
    \begin{tabular}{lc|lc} \toprule
    Action & Occurrence & Action & Occurrence \\ \midrule
    \texttt{Add}             & $744,533$ & \texttt{Wait}            & $38,211$\\
    \texttt{Stir}            & $287,413$ & \texttt{Recrystal.}   & $25,600$\\
    \texttt{Concentrate}     & $276,551$ & \texttt{PhaseSepa.} & $24,141$\\
    \texttt{Yield}           & $274,439$ & \texttt{PH}              & $21,756$  \\
    \texttt{MakeSolution}    & $272,537$ & \texttt{Quench}          & $18,699$  \\
    \texttt{Filter}          & $247,625$ & \texttt{Partition}       & $16,045$  \\
    \texttt{Wash}            & $224,286$ & \texttt{Triturate}       & $13,390$  \\
    \texttt{DrySolution}     & $178,248$ & \texttt{DrySolid}        & $6,435$   \\
    \texttt{CollectLayer}    & $146,379$ & \texttt{Degas}           & $4,789$   \\
    \texttt{Extract}         & $114,855$ & \texttt{Microwave}       & $2,237$   \\
    \texttt{SetTemp.}  & $44,126$  & \texttt{Sonicate}        & $450$     \\
    \texttt{Reflux}          & $43,296$  \\ \bottomrule
    \end{tabular}
    \caption{Action space and actions' occurrences in the OpenExp dataset.}
    \label{tab:action_count}
\end{table}
\begin{table*}[t]
\centering
\small
\begin{tabular}{ll} \toprule
\textbf{Field}                   & \textbf{Value}   \\
\midrule
Reactant                & \$1\$:  OC(CCc1ccccn1)C(F)(F)F                          \\
                        & \$3\$:  CC(C)(C){[}Si{]}(C)(C)Cl                        \\
                        & \$4\$:  c1c{[}nH{]}cn1                                 \\ \midrule
Solvent                 & \$2\$:  ClCCl                                           \\\midrule
Catalyst                & \$5\$:  CN(C)c1ccncc1                                   \\ \midrule
Product                 & \$-1\$:    CC(C)(C){[}Si{]}(C)(C)OC(CCc1ccccn1)C(F)(F)F \\
\midrule
Experimental & \textcolor{sky_blue}{\textbf{MAKESOLUTION}} with \$1\$ and \$2\$   (10 mL) ;           \\
Procedures              & \textcolor{sky_blue}{\textbf{ADD}}   \$3\$ (616 mg, 4.1 mmol, 1.2 eq) at 0°C ;       \\
                        & \textcolor{sky_blue}{\textbf{ADD}}   \$4\$ (697 mg, 10.2 mmol, 3.0 eq) at 0°C ;      \\
                        & \textcolor{sky_blue}{\textbf{ADD}}   \$5\$ (415 ng, 3.4 mmol) at 0°C ;               \\
                        & \textcolor{sky_blue}{\textbf{STIR}}   for 36 hours ;                                      \\
                        & \textcolor{sky_blue}{\textbf{CONCENTRATE}}   ;                                       \\
                        & \textcolor{sky_blue}{\textbf{YIELD}}   \$-1\$ (970 mg, 89\%). \\
\midrule
Source  & A solution of 700 mg (3.4 mmol) of 1,1,1-trifluoro-4-pyridin-2-ylbutan-2-ol in 10 mL \\
        & of dichloromethane was treated with 616 mg (4.1 mmol, 1.2 eq.) of \\
        & tert-butyldimethylsilyl chloride, 697 mg (10.2 mmol, 3.0 eq.) of imidazole and \\
        & 415 ng (3.4 mmol) of 4-dimethylaminopyridine at 0° C. The resulting mixture was \\
        & allowed to warm to room temperature and as stirred for 36 hours. Then the mixture \\
        & w was concentrated and the residue was purified by flash chromatography to give \\
        &  970 mg (89\%) of 2-{[}3-(tert-butyldimethylsilanyloxy)-4,4,4-trifluorobutyl{]}pyridine \\
        & as a colorless oil.
    \\\bottomrule
\end{tabular}
\caption{Illustrative example of the OpenExp dataset. \textcolor{sky_blue}{\textbf{BOLDED BLUE}} indicates pre-defined action.} 
\label{tab:OpenExp_example}
\end{table*}
\textbf{Preprocessing.} Following~\cite{smiles2actions}, we conduct preprocessing after the paragraph2action conversion, 
The preprocessing has two purposes: 1) extracting the important entities (\ie molecules) in experimental procedures and mapping all molecules to their precursors in the chemical reaction; 2) applying a rule-based filtration to improve the dataset quality. Our preprocessing strategy is inspired by~\cite{paragraph2actions}, augmented with additional 2 steps: perplexity filtering and similar action aggregation. The complete preprocessing steps are listed below:
\begin{itemize}[leftmargin=*]
    \item Perplexity Filtering. To ensure the quality of the above translation step, we compute a perplexity score for each output and exclude samples with a score larger than $1.0$. These perplexity scores are calculated using the TextChemT5 model.
    \item Entity Recognition. We extract all the molecules (either by name or SMILES) from the action sequences using the source codes of~\cite{paragraph2actions}.
    Then, we conduct string matching of IUPAC names between the extracted molecules and those in the chemical reactions. STOUT~\cite{stout} and PubChemPy\footnote{\url{https://github.com/mcs07/PubChemPy}} are used for the translation between IUPAC names and SMILES. If any molecule cannot be matched with its counterpart in the chemical reactions, we consider the reaction data invalid and remove it from the dataset. However, we permit the inclusion of certain common substances, such as common organic solvents, in every reaction. The names and SMILES expressions of the 134 common substances are included in our code. After entity recognition, we assign each entity a unique ID and update the experimental procedures by replacing the entity mentions with the corresponding entity IDs.
    \item Common Substance Renaming. We standardized the nomenclature for common substances that are known by multiple names (\eg water may also be referred to as H2O, pure water, water (aq.), \etc) to improve the dataset's precision. Using PubChemPy, we align the different names to their standardized SMILES representations, allowing us to identify when different terms refer to the same molecule by comparing their SMILES expressions.
    \item Similar Action Aggregation. If two adjacent operations are highly similar (\eg \textit{STIR} and \textit{STIR for 5 min}), they are merged together.
    \item Ensuring Single Product. This dataset focuses on the preparation of a single material, hence we remove reactions that yield multiple products.
    \item Action Filtering. We remove action sequences that have fewer than five actions or contain invalid actions.
    \item Reaction Deduplication. We remove the duplicated reactions from the dataset.
\end{itemize}

\begin{table}[t]
\centering
\small
\begin{tabular}{p{4.3cm}ll} \toprule
Total reactions                                             & 2262637 & 100\%   \\ \midrule
Too large perplexity score                                 & 329160  & 14.55\% \\
More than one product                                      & 105577  & 4.67\%  \\
Incomplete mapping of molecules (from chemical reaction)  & 1034908 & 45.74\% \\
Incomplete mapping of molecules (from action sequence)   & 178689  & 7.90\%  \\
Remove duplicate reactions                                 & 254099  & 11.23\% \\
Filter out too short actions                               & 14022   & 0.62\%  \\
Other errors                                               & 71743   & 3.16\%  \\ \midrule
Remaining reactions                                        & 274439  & 12.13\% \\ \bottomrule
\end{tabular}
\caption{Number of samples removed at each preprocessing step.}
\label{tab:remove_count}
\end{table}

Table \ref{tab:remove_count} presents the number of samples removed at each preprocessing step. Further, Table \ref{tab:OpenExp_example} provides an example from the final OpenExp dataset, we can observe that it encompasses:
\begin{itemize}[leftmargin=*]
    \item Structured, step-by-step instructions of experimental procedures; 
    \item All molecules in the reaction and their roles (\ie reactant, solvent, catalyst, product).
    \item The mapping between the recognized entities (\ie molecules) and their IDs. 
    \item The original unstructured experimental procedures.
\end{itemize}

\textbf{Discussion on License.} The ORD database is accessible under the CC-BY-SA license, and the USPTO-Applications dataset is available under the CC0 license. We have used codes from TextChemT5~\cite{TextChemT5} and Paragraph2Actions~\cite{smiles2actions}, which are both licensed under the MIT license. Therefore, we will release OpenExp under the CC-BY-SA license to comply with the most restrictive license of these resources. This license permits content distribution and sharing, provided the same license is applied.

\textbf{Human Evaluation.} \shiyr{We invite two PhD students majoring in chemistry to evaluate the quality of the OpenExp dataset. Specifically, 250 data points are randomly sampled from the dataset, and assigned to the evaluators according to the following rules: 1) the first 50 data points are assigned to both volunteers simultaneously to verify the consistency of their evaluations; 2) the remaining 200 data points are then evenly assigned to the two evaluators. Under this allocation rule, each evaluator is responsible for 150 data points.} Tthe evaluators are then asked to rate the quality of each data point on a scale from 1 (lowest) to 5 (highest). Our instructions to the evaluators are shown below:

\begin{tcolorbox}[title = {Instructions to human evaluators.}]
    \small
    We are curating a dataset partially generated by an AI model and want to seek feedback on its quality from human experts. During the evaluation process, we will provide both machine language sequences (the machine-generated operational sequences of experimental actions) and the corresponding natural language sequences (descriptions of experimental procedures in their original free texts).

    You should rate these samples based on how well the operational sequences align with the original descriptions. Please use a rating scale of 1 (low alignment) to 5 (high alignment). Molecular skeletal formulas are provided as images for reference during evaluation. All original data for this dataset come from the United States Patent and Trademark Office (USPTO), ensuring the viability of the reactions.
    \tcblower
    \small
    The following are the detailed scoring guidelines, with a maximum score of 5:
    \begin{itemize}[leftmargin=*, nosep]
        \item \textbf{5}: The machine-generated action sequence includes no errors in capturing key details of the original experimental procedure, including actions, materials, and numerical values.
        \item \textbf{4}: The machine-generated action sequence includes at most one ($n_{err} \leq 1$) error or omission related to actions, materials, or numerical values.
        \item \textbf{3}: The machine-generated action sequence includes at most two ($n_{err} \leq 2$) errors or omissions related to actions, materials, or numerical values.
        \item \textbf{2}: The machine-generated action sequence includes at most four ($n_{err} \leq 4$) errors or omissions related to actions, materials, or numerical values.
        \item \textbf{1}: The machine-generated action sequence includes more than four ($n_{err} > 4$) errors or omissions related to actions, materials, or numerical values.
    \end{itemize}
\end{tcolorbox}

\begin{figure}[t]
    \centering
    \includegraphics[width=.9\linewidth]{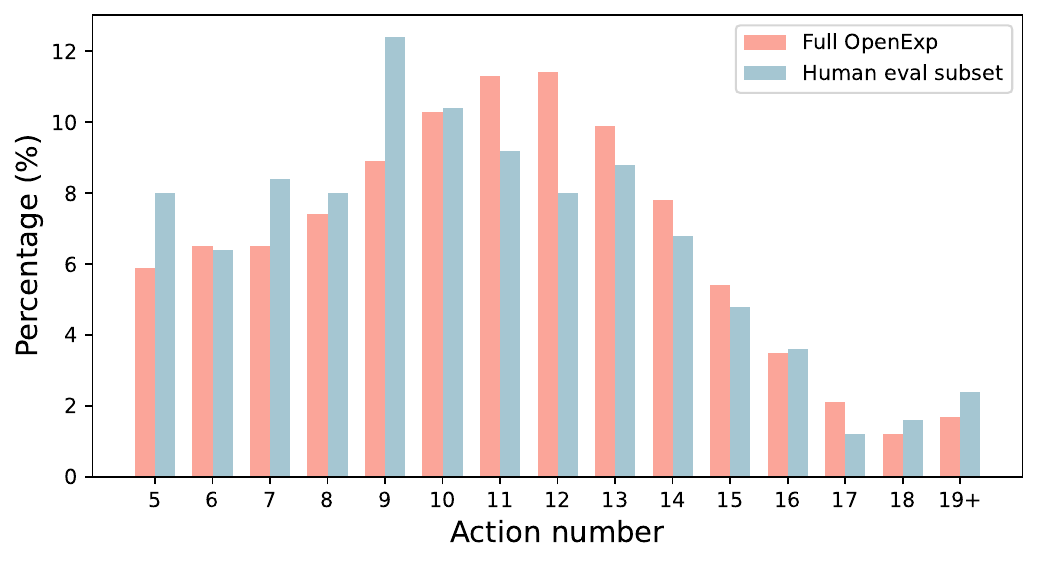}
    \caption{Action number distributions of the full OpenExp dataset and the human evaluation subset.}
    \label{fig:human_eval_distributions}
\end{figure}

\shiyr{Figure \ref{fig:human_evaluation} presents the human evaluation results. Statistics of these 250 data points and the entire dataset can be found in Figure~\ref{fig:human_eval_distributions} and Table~\ref{tab:human_eval_statistics}. We can observe that the distribution of the sampled data points closely resembles that of the entire dataset, suggesting that the human evaluation results can reflect the overall quality of the OpenExp dataset.}

\begin{figure}[t]
    \centering
    \includegraphics[width=.9\linewidth]{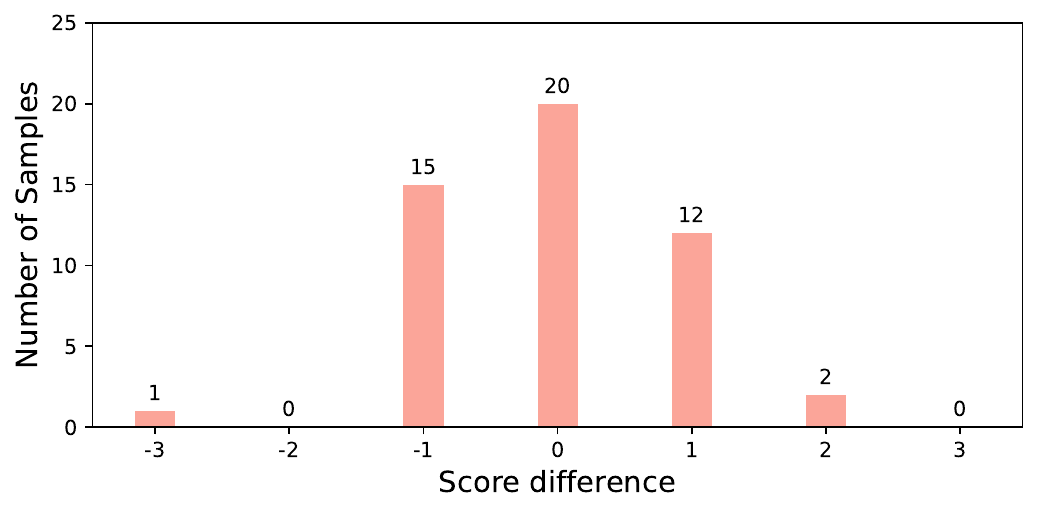}
    \caption{The score difference between evaluator 1 and evaluator 2 on 50 samples.}
    \label{fig:score_diff_50}
\end{figure}

\shiyr{Based on the 50 shared data points, we calculate the score differences in scores for the same samples (\ie, the score given by evaluator 1 minus the score given by evaluator 2). The results are presented in Figure~\ref{fig:score_diff_50}. We can observe the exact alignment in ratings for 40\% of the samples (20 out of 50), and a marginal score difference (±1) is recorded for 54\% of the samples (27 out of 50). Discrepancies of two or more scores are exceedingly rare, occurring in just 6\% of the samples (3 out of 50). Some examples of human evaluated data points are in Appendix~\ref{sec:app-human_cases}.}

\begin{table}[t]
    \centering
    \small
    \begin{tabular}{p{3.cm}ll}
        \toprule
        Property Name & Full OpenExp & Human eval. \\ \midrule
        Avg. Action Number   & 10.88        & 10.53      \\
        Avg. Reactant Number & 2.96         & 2.98       \\
        Avg. Product Number  & 1.00         & 1.00       \\
        Avg. Catalyst Number & 0.15         & 0.24       \\
        Avg. Solvent Number  & 1.06         & 1.00       \\
        Avg. Molecule Weight & 164.88       & 168.80     \\
        Avg. Atom Number     & 10.57        & 10.69      \\
        Avg. Bond Number     & 10.63        & 10.75      \\
        Avg. Ring Number     & 1.06         & 1.07       \\ 
        \bottomrule
    \end{tabular}
    \caption{Chemical property statistics of the full OpenExp dataset and the human evaluation subset. Human eval stands for the human evaluation subset.}
    \label{tab:human_eval_statistics}
\end{table}

\begin{table*}[t]
\centering
\small
\begin{tabular}{p{2.5cm}cp{2.2cm}cp{2cm}cp{1.8cm}c} \toprule
\multicolumn{2}{c}{\textbf{Computed Properties}}  & \multicolumn{6}{c}{\textbf{Experimental Properties}} \\ \cmidrule(lr){1-2} \cmidrule(lr){3-8}
\textbf{Property} &     \textbf{Count} &     \textbf{Property} &     \textbf{Count} &     \textbf{Property} &     \textbf{Count} &     \textbf{Property} &     \multicolumn{1}{c}{\textbf{Count}} \\ \cmidrule(lr){1-2} \cmidrule(lr){3-4} \cmidrule(lr){5-6} \cmidrule(lr){7-8}
Molecular Weight                    & 1244109 & Physical Description          & 8368 & Vapor Density             & 1043 & Enthalpy of Sublimation & 9 \\
Hydrogen Bond Donor Count           & 1244109 & Kovats Retention Index        & 6878 & Autoignition Temperature  & 771  & Acid Value              & 4 \\
Hydrogen Bond Acceptor Count        & 1244109 & Solubility                    & 5909 & Heat of Vaporization      & 583  & Dielectric Constant     & 2 \\
Rotatable Bond Count                & 1244109 & Chemical Classes              & 5726 & Viscosity                 & 550  & Dispersion              & 1 \\
Exact Mass                          & 1244109 & Melting Point                 & 4468 & Taste                     & 514  & Hydrophobicity          & 1 \\
Monoisotopic Mass                   & 1244109 & Vapor Pressure                & 3032 & Henry's Law Constant      & 502  &                         &   \\
Topological Polar Surface Area      & 1244109 & Boiling Point                 & 2996 & Surface Tension           & 448  &                         &   \\
Heavy Atom Count                    & 1244109 & Color/Form                    & 2927 & pH                        & 444  &                         &   \\
Formal Charge                       & 1244109 & Density                       & 2862 & Odor Threshold            & 442  &                         &   \\
Complexity                          & 1244109 & LogP                          & 2763 & Corrosivity               & 410  &                         &   \\
Isotope Atom Count                  & 1244109 & Other Experimental Properties & 2393 & Heat of Combustion        & 405  &                         &   \\
Defined Atom Stereocenter   Count   & 1244109 & Decomposition                 & 2033 & Ionization Efficiency     & 332  &                         &   \\
Undefined Atom Stereocenter   Count & 1244109 & Refractive Index              & 1777 & Optical Rotation          & 265  &                         &   \\
Defined Bond Stereocenter   Count   & 1244109 & Collision Cross Section       & 1634 & Ionization Potential      & 253  &                         &   \\
Undefined Bond Stereocenter   Count & 1244109 & Odor                          & 1512 & LogS                      & 166  &                         &   \\
Covalently-Bonded Unit Count        & 1244109 & Stability/Shelf Life          & 1506 & Polymerization            & 134  &                         &   \\
Compound Is Canonicalized           & 1244109 & Flash Point                   & 1479 & Relative Evaporation Rate & 101  &                         &   \\
XLogP3                              & 1184175 & Dissociation Constants        & 1250 & Caco2 Permeability        & 79   &                         &  \\\bottomrule
\end{tabular}
\vspace{-3mm}
\caption{Statistics of the collected molecule properties, including computed properties and experimental properties.}
\vspace{-3mm}
\label{tab:molecule_properties}
\end{table*}

\subsection{Collection and Preprocessing of ReactXT's Pretraining Dataset}
\label{sec:app-pretrain_dataset}

In Section~\ref{sec:pretrain}, we collect and compile a dataset to incrementally pretrain an LM for improved understanding of chemical reactions and individual molecules. Here we elaborate on the details of this dataset, which includes the following contents:
\begin{itemize}[leftmargin=*]
    \item A total of 1,162,551 chemical reactions;
    \item Patent abstracts and computed/experimental properties of 1,254,157 molecules, which are all from the chemical reactions.
\end{itemize}

We extract chemical reactions from ORD and USPTO datasets. Then, we source patent abstracts from PubChem's Patent View\footnote{\url{pubchem.ncbi.nlm.nih.gov/docs/patents}} and obtain molecular properties using the PubChem's PubView API\footnote{\url{pubchem.ncbi.nlm.nih.gov/docs/pug-view}}. For each molecule, the abstract text derives from the abstracts of patent documents where the molecule is mentioned, and its properties include both computational and experimental ones. Table \ref{tab:molecule_properties} shows a complete list of these properties.

In Table \ref{tab:pretrain_statistics}, we compare the statistics of our pretraining dataset with that of PubChem324k. We can observe that ReactXT's pretraining dataset includes more molecules and additionally includes chemical reactions.

\begin{table}[t]
    \centering
    \small
    \begin{tabular}{l|cc} \toprule
        & Our Dataset & Pubchem324k \\ \midrule
        Num of Molecules         & $1,254,157$ & $313,083$ \\
        Num of Reactions         & $1,162,551$ & -         \\
        Avg. Molecule Weight     & $362.4$     & $502.4$   \\
        Avg. Atom Count         & $24.9$      & $35.2$    \\
        Avg. Bond Count         & $26.8$      & $37.6$    \\
        Avg. Ring Count         & $2.9$     & $3.5$   \\
        Avg. Text Length         & $517.8$     & $120.4$   \\
        Avg. Property Count      & $17.8$      & -         \\ \bottomrule
    \end{tabular}
    \vspace{-3mm}
    \caption{Statistics of ReactXT's pretraining dataset and Pubchem324k.}
    \vspace{-3mm}
    \label{tab:pretrain_statistics}
\end{table}

To prevent information leakage, we exclude a total of 54,403 reactions that appear in the validation and test sets of the downstream datasets (\ie OpenExp and USPTO-50K~\cite{uspto-50k}) from the pretraining dataset. The remaining 1,108,148 reactions are used for pretraining.

\textbf{Discussion on License.} The ORD database is accessible under the CC-BY-SA license, and the USPTO-Applications dataset is available under the CC0 license. The patent abstracts from PubChem are provided by Google Patent\footnote{\url{patents.google.com}}, which is released under the CC-BY-4.0 license. To comply with the strictest license terms, we will release our dataset under the CC-BY-SA license. 

Additionally, we have utilized textual descriptions, computed properties, and experimental properties from the PubChem website for pretraining. Given that this data is aggregated from various sources by PubChem, determining a single appropriate license is challenging. To support future research while avoiding licensing complexities, we will provide the scripts for downloading and preprocessing this data, rather than distributing the data directly.

\section{Experimental Details}
\label{sec:app-exp}

\subsection{Hyperparameters}
Here we detail the hyperparameters for ReactXT's pretraining and finetuning across three downstream tasks. 
Due to the prohibitive costs associated with training large LMs, finetuning on downstream datasets is limited to a single run.


\begin{table*}[t]
    \small
    \centering
    \setlength{\tabcolsep}{2pt}
    \resizebox{0.9\linewidth}{!}{
    \begin{tabular}{ll|cccccc} \toprule
    \textbf{Pretrain   Input Context} &
        \textbf{Pretrain Data Type} &
        \textbf{BLEU-2} &
        \textbf{BLEU-4} &
        \textbf{ROUGE-1} &
        \textbf{ROUGE-2} &
        \textbf{ROUGE-L} &
        \textbf{METEOR} \\ \midrule
    No incremental pretrain & -        & 39.4 & 32.2 & 52.7 & 39.4 & 47.6 & 49.2 \\ 
    Reactions               & reaction & 37.3 & 29.9 & 50.3 & 36.5 & 45.0 & 46.7 \\
    ReactXT &
        reaction, sing. mol. &
        \textbf{42.6} &
        \textbf{35.2} &
        \textbf{54.7} &
        \textbf{41.7} &
        \textbf{49.6} &
        \textbf{51.2} \\ \bottomrule
    \end{tabular}}
    \caption{Ablation study. Performances (\%) for molecule captioning on the PubChem324k dataset.}
    \label{tab:ablate_caption}
    \end{table*}
\textbf{ReactXT Pretrain.}
The pretraining stage of ReactXT has 5 million steps, with the number of molecules per reaction being $k=4$. Following MolCA's~\cite{MolCA} experimental setup, we employ a Q-former with 8 query tokens. We use AdamW as the optimizer, with a weight decay set to $0.05$. The optimizer's peak learning rate is set to $1 \times 10^{-4}$, scheduled by linear warmup with cosine decay. The warmup has 1000 steps and starts at a learning rate of $1 \times 10^{-6}$.

\textbf{Experimental Procedure Prediction.}
We fully finetune all the baseline methods and ReactXT for $20$ epochs, with a batch size of $32$. The optimizer and learning rate settings are consistent with the pretraining phase.

\shiyr{\textbf{Retrosynthesis.} Following~\cite{r-smiles}, we sample 20 root-aligned augmentations for the training and testing subsets. Before finetuning on USPTO-50K, We first conduct 2 epochs of masked self-supervised pretraining for MolT5 and ReactXT on the USPTO-full dataset \cite{GLN}, following the pretraining strategy of R-SMILES \cite{r-smiles}. During finetuning, we train MolT5 for 20 epochs and ReactXT for 5 epochs on the augmented training set using a batch size of 32. We then average the model's parameters on the last several tuning steps as the final checkpoint for testing. During testing, we conduct a beam search with a beam size of 20 for both models and return the top ten results as the model's predictions. The beam size (20) and the number of results (10) are following the experiment of R-SMILES \cite{r-smiles}. The optimizer and learning rate settings are kept consistent with the pretraining phase.}


\textbf{Molecule Captioning.} On both datasets, we full finetune MolCA and ReactXT $20$ epochs, with a batch size of $32$. The optimizer and learning rate settings are consistent with the pretraining phase.

\subsection{Other Implementation Details}

\textbf{Baselines.} We briefly introduce the baselines:
\begin{itemize}[leftmargin=*]
    \item \textbf{Galactica}~\cite{Galactica}. Galactica is a scientific language model which is pretrained on 2 million compounds from PubChem. It has a decent understanding of SMILES formulas. 
    \item \textbf{MolT5}~\cite{MolT5}. MolT5 is developed based on the T5 model. Its training corpora include both natural language and SMILES data, making it suitable for both molecule captioning and text-based molecular generation tasks.
    \item \textbf{TextChemT5}~\cite{TextChemT5}. TextChemT5 is a T5-based multi-domain LM, which is tuned on various text-molecule tasks.
    \item \textbf{MolCA}~\cite{MolCA}. MolCA is a multimodal language model finetuned on Galactica. It includes both graph encoder and LM, where a Querying Transformer is applied to align their latent spaces. 
    \item \textbf{AT}~\cite{AT}. AT trains transformers with data augmentation for retrosynthesis. The data augmentation is achieved by rearranging the order of characters in SMILES strings in both the training and test sets.
    \item \textbf{MEGAN}~\cite{megan}. MEGAN represents chemical reactions as a sequence of graph edits and performs retrosynthesis by sequentially modifying the target molecule.
    \item \textbf{MoMu}~\cite{MoMu}. Momu contrastively pretrains a GNN and an LM with paired molecular graph-text data, and can be adapted to retrieval and generation tasks.
    \item \textbf{Chemformer}~\cite{chemformer}. Chemformer is a Transformer-based molecule LM that is self-supervised pretrained on a SMILES corpus. It can be applied to both generation and property prediction tasks.
    \item \textbf{Random, among all reactions}~\cite{smiles2actions}. Randomly pick an action sequence from the training set.
    \item \textbf{Random, compatible pattern}~\cite{smiles2actions}. Randomly pick an action sequence from the training subset of reactions that have the same number of molecules as the current reaction.
    \item \textbf{Nearest Neighbor}~\cite{smiles2actions}. Pick the action sequence from the training set with the reaction most similar to the current one, as determined by reaction fingerprints~\cite{schwaller2019data}.
\end{itemize}

\begin{table}[t]
    \centering
    \small
    \setlength{\tabcolsep}{2pt}
    \resizebox{\linewidth}{!}{
        \begin{tabular}{l|ccccc}
            \toprule
            & BLEU2                     & BLEU4                     & ROUGE1                    & ROUGE2                    & ROUGEL                    \\
            \midrule
            T-statistic & 14.619                    & 13.622                    & 16.126                    & 14.438                    & 15.053                    \\
            P-value     & \textbf{<0.001} & \textbf{<0.001} & \textbf{<0.001} & \textbf{<0.001} & \textbf{<0.001} \\
            \bottomrule
        \end{tabular}
    }
    \caption{P-values for experimental procedure prediction (Table~\ref{tab:exp1_openexp}), comparing ReactXT against MolCA-1.3B.}
    \label{tab:t_test_action}
\end{table}

\begin{table}[t]
    \centering
    \small
    \setlength{\tabcolsep}{2pt}
    \resizebox{\linewidth}{!}{
        \begin{tabular}{l|cccccc}
            \toprule
            &
             BLEU2 & BLEU4 & ROUGE1 & ROUGE2 & ROUGEL & METEOR \\
            \midrule
            T-statistic & 3.469 & 3.823 & 3.451 & 3.851 & 3.434 & 4.107 \\
           P-value & \textbf{<0.001} & \textbf{<0.001} & \textbf{<0.001} & \textbf{<0.001} & \textbf{<0.001} & \textbf{<0.001} \\
            \bottomrule
        \end{tabular}
    }
    \caption{P-values for captioning on PubChem324k (Table~\ref{tab:caption}), comparing ReactXT against MolCA-1.3B, full ft. }
    \label{tab:t_test_caption_a}
\end{table}

\begin{table}[t]
    \centering
    \small
    \setlength{\tabcolsep}{2pt}
    \resizebox{\linewidth}{!}{
        \begin{tabular}{l|cccccc}
            \toprule
            & BLEU2          & BLEU4                     & ROUGE1         & ROUGE2                    & ROUGEL         & METEOR \\
            \midrule
            T-statistic & 2.918          & 3.523                     & 2.843          & 3.495                     & 3.129          & 2.195                      \\
            P-value     & \textbf{0.004} & \textbf{<0.001} & \textbf{0.004} & \textbf{<0.001} & \textbf{0.002} & \textbf{0.028} \\     
            \bottomrule
        \end{tabular}
    }
    \caption{P-values for captioning on CheBI-20 (Table~\ref{tab:caption}), comparing ReactXT against MolCA-1.3B, full ft. }
    \label{tab:t_test_caption_b}
\end{table}

\begin{table}[t]
    \centering
    \small
    \setlength{\tabcolsep}{2pt}
        \begin{tabular}{l|cccc}
        \toprule
        & Top1           & Top3           & Top5  & Top10 \\
        \midrule
        Z-statistic & 2.340          & 2.380          & 0.440 & 0.000 \\
        P-value     & \textbf{0.019} & \textbf{0.017} & 0.662 & 1.000 \\
        \bottomrule
        \end{tabular}
    \caption{P-values for retrosynthesis (Table~\ref{tab:retrosynthesis}), comparing ReactXT against R-SMILES. Both models use 20 augmentations during testing.}
    \label{tab:t_test_retro}
\end{table}

\begin{table*}[]
\small
\centering
\begin{subtable}{\textwidth}
\begin{tabular}{p{1.5cm}l|p{1.5cm}l} \toprule
\textbf{Field} & \multicolumn{3}{l}{\textbf{Value}} \\ \midrule
Reactant & \multicolumn{3}{l}{\$1\$:  OCCCCCCCc1ccccc1} \\
    & \multicolumn{3}{l}{\$2\$:  C\#CC(=O)O} \\
    & \multicolumn{3}{l}{\$4\$:  c1ccccc1} \\ \midrule
Catalyst & \multicolumn{3}{l}{\$3\$:  Cc1ccc(S(=O)(=O)O)cc1} \\ \midrule
Product & \multicolumn{3}{l}{\$-1\$:   C\#CC(=O)OCCCCCCCc1ccccc1} \\ \midrule
Source & \multicolumn{3}{p{13cm}}{A mixture of 0.5 g of   7-phenylheptanol, 0.27 g of propiolic acid, 0.005 g of p-toluenesulfonic acid   and 25 ml of benzene was refluxed with stirring for six hours while water   formed was removed by a Dean-Stark water separator. After the reaction was   completed, the reaction solution was washed successively with a 5\% aqueous   sodium bicarbonate solution and a saturated sodium chloride solution, and   dried over anhydrous magnesium sulfate. After removal of the solvent under   reduced pressure, the obtained residue was subjected to silicagel column   chromatography to yield 0.368 g of 7-phenylheptyl propiolate (compound 3).} \\ \midrule
Annotated Actions & \begin{tabular}[c]{@{}p{5.5cm}@{}}\textcolor{sky_blue}{\textbf{MAKESOLUTION}} with \$1\$ (1.1 g)   and \$2\$ (0.005 g) and \$3\$ (25 ml) and \$4\$ ; \\ \textcolor{sky_blue}{\textbf{REFLUX}} for 6 hours ; \\ \textcolor{sky_blue}{\textbf{CONCENTRATE}} ;\\ \textcolor{sky_blue}{\textbf{WASH}} with NaHCO3 ;\\ \textcolor{sky_blue}{\textbf{WASH}} with sodium chloride ; \\ \textcolor{sky_blue}{\textbf{DRYSOLUTION}} over magnesium sulfate ; \\ \textcolor{sky_blue}{\textbf{FILTER}} keep filtrate ;\\ \textcolor{sky_blue}{\textbf{YIELD}} \$-1\$ (1.15 g).\end{tabular} & Predicted Actions & \begin{tabular}[c]{@{}p{5.5cm}@{}}\textcolor{sky_blue}{\textbf{MAKESOLUTION}} with \$1\$ (0.27 g)   and \$2\$ (0.005 g) and \$3\$ (25 ml) and \$4\$ ; \\      \textcolor{sky_blue}{\textbf{REFLUX}} for 10 hours; \\      \textcolor{sky_blue}{\textbf{CONCENTRATE}} ; \\      \textcolor{sky_blue}{\textbf{WASH}} with NaHCO3 ;\\      \textcolor{sky_blue}{\textbf{WASH}} with sodium chloride ;\\      \textcolor{sky_blue}{\textbf{DRYSOLUTION}} over magnesium sulfate ;\\      \textcolor{sky_blue}{\textbf{FILTER}} keep filtrate ; \\      \textcolor{sky_blue}{\textbf{YIELD}} \$-1\$ (0.368 g).\end{tabular} \\\bottomrule
\end{tabular}
\caption{Example 1.}
\end{subtable}
\begin{subtable}{\textwidth}
    \begin{tabular}{p{1.5cm}l|p{1.5cm}l}  \toprule
        \textbf{Field}    & \multicolumn{3}{l}{\textbf{Value}} \\ \midrule
        Reactant & \multicolumn{3}{l}{\$1\$:    C{[}Si{]}1(C)CC{[}Si{]}(C)(C)N1c1ccc(C(O)c2cn(S(=O)(=O)c3ccccc3)c3ncc(Cl)cc23)cn} \\
                    & \multicolumn{3}{l}{\$2\$:    Nc1ccc(C(O)c2cn(S(=O)(=O)c3ccccc3)c3ncc(Cl)cc23)cn1}                              \\
                    & \multicolumn{3}{l}{\$4\$:  CC{[}SiH{]}(CC)CC}                                                                  \\
                    & \multicolumn{3}{l}{\$5\$:    O=C(O)C(F)(F)F}                                                                   \\\midrule
        Solvent  & \multicolumn{3}{l}{\$3\$:  ClCCl} \\ \midrule
        Product  & \multicolumn{3}{l}{\$-1\$:   Nc1ccc(Cc2cn(S(=O)(=O)c3ccccc3)c3ncc(Cl)cc23)cn1}                                 \\ \midrule
        Source &
            \multicolumn{3}{p{13cm}}{To   (1-benzenesulfonyl-5-chloro-1H-pyrrolo{[}2,3-b{]}pyridin-3-yl)-{[}6-(2,2,5,5-tetramethyl-{[}1,2,5{]}azadisi-lolidin-1-yl)-pyridin-3-yl{]}-methanol   and   (6-amino-pyridin-3-yl)-(1-benzenesulfonyl-5-chloro-1H-pyrrolo{[}2,3-b{]}pyridin-3-yl)-methanol   (118, 119, 1.70/1.25 g mix, 2.41 mmol) in 25.0 mL of dichloromethane,   triethylsilane (3.00 mL, 18.8 mmol) and trifluoroacetic acid (1.50 mL, 19.5   mmol) were added and the reaction stirred at room temperature overnight. The   reaction was concentrated under vacuum, combined with aqueous potassium   carbonate and extracted with ethyl acetate. The organic layer was dried over   sodium sulfate, filtered and the filtrate concentrated under vacuum. The   resulting material was purified by silica gel column chromatography eluting   with 20-100\% ethyl acetate in hexane to provide the desired compound (120,   0.70 g).} \\ \midrule
        Annotated Actions &
            \begin{tabular}[c]{@{}p{5.5cm}@{}} \textcolor{sky_blue}{\textbf{MAKESOLUTION}} with \$1\$ and \$2\$   and \$3\$ (25.0 mL) ;\\      \textcolor{sky_blue}{\textbf{ADD}} \$4\$ (3.00 mL, 18.8 mmol) ;\\      \textcolor{sky_blue}{\textbf{ADD}} \$5\$ (1.50 mL, 19.5 mmol) ;\\      \textcolor{sky_blue}{\textbf{STIR}} for overnight at room temperature ;\\      \textcolor{sky_blue}{\textbf{CONCENTRATE}} ;\\      \textcolor{sky_blue}{\textbf{ADD}} K2CO3 ;\\      \textcolor{sky_blue}{\textbf{EXTRACT}} with ethyl acetate ;\\      \textcolor{sky_blue}{\textbf{COLLECTLAYER}} organic ;\\      \textcolor{sky_blue}{\textbf{DRYSOLUTION}} over sodium sulfate ;\\      \textcolor{sky_blue}{\textbf{FILTER}} keep filtrate ;\\      \textcolor{sky_blue}{\textbf{CONCENTRATE}} ;\\      \textcolor{sky_blue}{\textbf{YIELD}} \$-1\$ (0.70 g).\end{tabular} & Predicted Actions &
            \begin{tabular}[c]{@{}p{6cm}@{}} \textcolor{sky_blue}{\textbf{MAKESOLUTION}} with \$1\$ (1.00 g,   1.91 mmol) and \$2\$ (0.69 g, 1.72 mmol) and \$3\$ (35 mL) ;\\      \textcolor{sky_blue}{\textbf{ADD}} \$4\$ (1.35 mL, 7.84 mmol) ;\\      \textcolor{sky_blue}{\textbf{ADD}} \$5\$ (1.90 mL, 26.7 mmol) ;\\      \textcolor{sky_blue}{\textbf{STIR}} for 8 h at room temperature ;\\      \textcolor{sky_blue}{\textbf{CONCENTRATE}} ;\\      \textcolor{sky_blue}{\textbf{EXTRACT}} with K2CO3 ;\\      \textcolor{sky_blue}{\textbf{EXTRACT}} with ethyl acetate ; \\      \textcolor{sky_blue}{\textbf{COLLECTLAYER}} organic ;\\      \textcolor{sky_blue}{\textbf{DRYSOLUTION}} over sodium sulfate ; \\      \textcolor{sky_blue}{\textbf{FILTER}} keep filtrate ;\\      \textcolor{sky_blue}{\textbf{CONCENTRATE}} ;\\      \textcolor{sky_blue}{\textbf{YIELD}} \$-1\$ (0.13 g, 19\%).\end{tabular} \\ \bottomrule
        \end{tabular}
    \caption{Example 2.}
\end{subtable}
\caption{Examples of accurate experimental procedure predictions.}
\label{tab:case_12}
\end{table*}

\begin{table*}[t]
\centering
\small
\begin{subtable}{\textwidth} 
\begin{tabular}{p{1.5cm}l|p{1.5cm}l} \toprule
    \textbf{Field}       & \multicolumn{3}{l}{\textbf{Value}} \\\midrule
    Reactant             & \multicolumn{3}{l}{\$1\$:    Nc1ccc(C(=O)N{[}C@H{]}(CO)Cc2ccccc2)c(/C=C/c2ccccc2)c1} \\
                            & \multicolumn{3}{l}{\$3\$:  CC(=O)OC(C)=O} \\\midrule
    Solvent              & \multicolumn{3}{l}{\$2\$:  C1CCOC1} \\\midrule
    Product              & \multicolumn{3}{l}{\$-1\$:   CC(=O)Nc1ccc(C(=O)N{[}C@H{]}(CO)Cc2ccccc2)c(/C=C/c2ccccc2)c1} \\\midrule
    Source               & \multicolumn{3}{p{12cm}}{1 g (2.7 mmol) of   (S)-4-amino-2(E-2-phenylethen-1-yl)-N-(3-phenylpropan-1-ol-2-yl)benzamide   (intermediate 43f) was suspended in 50 ml of tetrahydrofuran and mixed with   0.25 ml (2.7 mmol) of acetic anhydride at 100° C. The mixture was stirred for   16 h. The reaction was then concentrated under reduced pressure and the   residue was recrystallized from ethanol. 0.78 g (71\%) of the product was   obtained.} \\ \midrule
    Annotated Actions & \begin{tabular}[c]{@{}p{4.5cm}@{}}\textcolor{sky_blue}{\textbf{ADD}} \$1\$ (1 g, 2.7 mmol) ;\\      \textcolor{sky_blue}{\textbf{ADD}} \$2\$ (50 ml) ;\\      \textcolor{sky_blue}{\textbf{ADD}} \$3\$ (0.25 ml, 2.7 mmol) ;\\      \textcolor{sky_blue}{\textbf{STIR}} for 16 h at 100 °C ;\\      \textcolor{sky_blue}{\textbf{CONCENTRATE}} ;\\      \textcolor{sky_blue}{\textbf{RECRYSTALLIZE}} from ethanol (0.78 g) ;\\      \textcolor{sky_blue}{\textbf{YIELD}} \$-1\$.\end{tabular} & Predicted Actions & \begin{tabular}[c]{@{}p{6cm}@{}} \textcolor{sky_blue}{\textbf{MAKESOLUTION}} with \$1\$ and \$2\$   (15 ml) ;\\      \textcolor{sky_blue}{\textbf{ADD}} \$3\$ (0.27 ml, 2.9 mmol) ;\\      \textcolor{mygreen}{\textbf{STIR} for 10 h at 100 °C ;}\\      \textcolor{mygreen}{\textbf{WAIT} for 10 h at 100 °C ;}\\      \textcolor{mygreen}{\textbf{STIR} for 10 h at 100 °C ;} \\    \textcolor{sky_blue}{\textbf{CONCENTRATE} ;}\\      \textcolor{sky_blue}{\textbf{ADD}} ethyl acetate ;    \\ \textcolor{sky_blue}{\textbf{ADD}} water ;\\      \textcolor{sky_blue}{\textbf{PHASESEPARATION}} ;\\      \textcolor{mygreen}{\textbf{COLLECTLAYER} aqueous ;}\\      \textcolor{sky_blue}{\textbf{EXTRACT}} with ethyl acetate ; \\     \textcolor{mygreen}{\textbf{COLLECTLAYER} organic ;}\\      \textcolor{sky_blue}{\textbf{WASH}} with brine ;\\      \textcolor{sky_blue}{\textbf{DRYSOLUTION}} over magnesium sulfate ;\\      \textcolor{sky_blue}{\textbf{FILTER}} keep filtrate ;\\      \textcolor{sky_blue}{\textbf{CONCENTRATE}} ; \\      \textcolor{sky_blue}{\textbf{YIELD}} \$-1\$ (144 mg, 75\%).\end{tabular} \\ \bottomrule
    \end{tabular}
\caption{Example 3.}
\end{subtable}
\begin{subtable}{\textwidth}
\begin{tabular}{p{1.5cm}l|p{1.5cm}l} \toprule
\textbf{Field}       & \multicolumn{3}{l}{\textbf{Value}} \\ \midrule
Reactant             & \multicolumn{3}{l}{\$1\$: Brc1ccc2noc(-c3ccccc3)c2c1} \\
                        & \multicolumn{3}{l}{\$2\$: O} \\ \midrule
CATALYST             & \multicolumn{3}{l}{\$3\$: {[}Zn{]}} \\ \midrule
SOLVENT              & \multicolumn{3}{l}{\$4\$: CC(=O)O} \\ \midrule
PRODUCT              & \multicolumn{3}{l}{\$-1\$: Nc1ccc(Br)cc1C(=O)c1ccccc1} \\ \midrule
Source               & \multicolumn{3}{p{13cm}}{5-Bromo-3-phenyl-2,1-benzisoxazole (7.5 g, 28.6 m   mol), water (14.6 ml), and zinc dust (9.3 g, 143 m mol) were combined. Acetic   acid (8.6 ml, 143 m mol) was added and the mixture was stirred and heated at   80° C. for 90 minutes. After cooling to room temperature, both the liquid and   solid portion of the reaction were extracted with methylene chloride. The   combined methylene chloride solutions was washed once with sodium hydroxide   solution (10\%) and several times with water. Drying (sodium sulfate) and evaporation   yielded the desired product (7.42 g) of mp. 92°-102° C.} \\ \midrule
Annotated Actions & \begin{tabular}[c]{@{}p{5cm}@{}} \textcolor{sky_blue}{\textbf{ADD}} \$1\$ (7.5 g) ;\\      \textcolor{sky_blue}{\textbf{ADD}} \$2\$ (14.6 ml) ;\\      \textcolor{sky_blue}{\textbf{ADD}} \$3\$ (9.3 g) ;\\      \textcolor{sky_blue}{\textbf{ADD}} \$4\$ (8.6 ml) ;\\      \textcolor{sky_blue}{\textbf{STIR}} for 90 minutes at 80 °C ;\\      
\textcolor{sky_blue}{\textbf{EXTRACT}} with methylene chloride ;\\      \textcolor{sky_blue}{\textbf{EXTRACT}} with CH2 Cl2 ;\\      \textcolor{sky_blue}{\textbf{WASH}} with NaOH ;\\      \textcolor{sky_blue}{\textbf{WASH}} with \$2\$ ;\\      \textcolor{sky_blue}{\textbf{DRYSOLUTION}} over sodium sulfate ;\\      \textcolor{sky_blue}{\textbf{FILTER}} keep filtrate ; \\ \textcolor{sky_blue}{\textbf{YIELD}} \$-1\$ (7.42 g).\end{tabular} & Predicted Actions & \begin{tabular}[c]{@{}p{6.5cm}@{}} \textcolor{sky_blue}{\textbf{MAKESOLUTION}} with \$1\$ and \$2\$   (150 ml) ;\\      \textcolor{sky_blue}{\textbf{ADD}} \$3\$ (24 g) ;\\      \textcolor{sky_blue}{\textbf{ADD}} \$4\$ (60 ml) ;\\      \textcolor{mygreen}{\textbf{STIR} for 120 minutes at 80 °C ;}\\      \textcolor{mygreen}{\textbf{WAIT} for 120 minutes at 80 °C ;}\\      \textcolor{sky_blue}{\textbf{FILTER}} ;\\      \textcolor{sky_blue}{\textbf{ADD}} carbon ;\\      \textcolor{sky_blue}{\textbf{COLLECTLAYER}} organic ;\\      \textcolor{sky_blue}{\textbf{WASH}} with \$2\$ (200 ml) 3 x ;\\      \textcolor{sky_blue}{\textbf{DRYSOLUTION}} over sodium sulfate ;\\      \textcolor{sky_blue}{\textbf{FILTER}} keep filtrate ;\\      \textcolor{sky_blue}{\textbf{CONCENTRATE}} ;\\      \textcolor{sky_blue}{\textbf{RECRYSTALLIZE}} from 2-amino-5-bromo-benzophenone (20.7 g) ;\\      \textcolor{sky_blue}{\textbf{YIELD}} \$-1\$ (20.7 g, 57.9\%).\end{tabular} \\\bottomrule
\end{tabular}
\caption{Example 4.}
\end{subtable}
\caption{Examples of inaccurate experimental procedure predictions. \textcolor{mygreen}{Green} denotes error of repetition.}
\label{tab:case_34}
\end{table*}

\begin{table*}[t]
\small
\centering
\begin{subtable}{\textwidth}
\begin{tabular}{p{1.5cm}l|p{1.5cm}l} \toprule
    \textbf{Field} & \multicolumn{3}{l}{\textbf{Value}}                                                                        \\ \midrule
    Reactant       & \multicolumn{3}{l}{\$1\$: COc1ccc(-c2cccc(CC(=O)O)c2)cc1C(C)=NO}                                            \\
                    & \multicolumn{3}{l}{\$3\$: FC(F)(F)c1ccc(CBr)cc1}                                                            \\
                    & \multicolumn{3}{l}{\$4\$: {[}H-{]}.{[}Na+{]}}                                                               \\
                    & \multicolumn{3}{l}{\$5\$: Cl}                                                                               \\ \midrule
    SOLVENT        & \multicolumn{3}{l}{\$2\$: C1CCOC1}                                                                          \\ \midrule
    PRODUCT        & \multicolumn{3}{l}{\$-1\$:   COc1ccc(-c2cccc(CC(=O)O)c2)cc1C(C)=NOCc1ccc(C(F)(F)F)cc1}                    \\ \midrule
    Source &
        \multicolumn{3}{p{13cm}}{9 mg of   {[}3'-(1-Hydroxyiminoethyl)-4'-methoxybiphenyl-3-yl{]}acetic acid was dissolved   in 0.2 mg of tetrahydrofuran, then 25 mg of 4-trifluoromethylbenzyl bromide   and 10 mg of sodium hydride were added thereto, and the mixture was stirred   at room temperature for 3 days. The reaction mixture was neutralized with 1N   hydrochloric acid and then extracted with ethyl acetate. The organic layer   was concentrated, and the residue was purified by reverse-phase high   performance liquid chromatography to give 5.21 mg of the title compound.} \\ \midrule
    Annotation   Actions &
        \begin{tabular}[c]{@{}p{4.8cm}@{}}\textcolor{sky_blue}{\textbf{ADD}} \$1\$ (9 mg) ;\\      \textcolor{sky_blue}{\textbf{ADD}} \$2\$ (0.2 mg) ;\\      \textcolor{sky_blue}{\textbf{ADD}} \$3\$ (25 mg) ;\\      \textcolor{sky_blue}{\textbf{ADD}} \$4\$ (10 mg) ;\\      \textcolor{sky_blue}{\textbf{STIR}} for 3 days at room temperature ;\\      \textcolor{sky_blue}{\textbf{PH}} with \$5\$ (1 normal) to pH neutral at room temperature ; \\      \textcolor{sky_blue}{\textbf{EXTRACT}} with ethyl acetate ;\\      \textcolor{sky_blue}{\textbf{CONCENTRATE}} ;\\      \textcolor{sky_blue}{\textbf{YIELD}} \$-1\$ (5.21 mg).\end{tabular} &
        Predicted Actions &
        \begin{tabular}[c]{@{}p{6.5cm}@{}} \textcolor{sky_blue}{\textbf{MAKESOLUTION}} with \$1\$ (25 mL)   and \$2\$ ;\\      \textcolor{sky_blue}{\textbf{ADD}} \$3\$ (292 mg) ;\\      \textcolor{sky_blue}{\textbf{ADD}} \$4\$ (104 mg) ;\\      \textcolor{sky_blue}{\textbf{STIR}} for overnight at room temperature ;\\      \textcolor{sky_blue}{\textbf{PH}} with \$5\$ (1 normal) to pH neutral at room temperature ;\\      \textcolor{sky_blue}{\textbf{COLLECTLAYER}} organic ;\\      \textcolor{sky_blue}{\textbf{WASH}} with ethyl acetate ; \\      \textcolor{sky_blue}{\textbf{COLLECTLAYER}} organic ;\\      \textcolor{sky_blue}{\textbf{WASH}} with water ;\\      \textcolor{sky_blue}{\textbf{WASH}} with brine ;\\      \textcolor{sky_blue}{\textbf{DRYSOLUTION}} over magnesium sulfate ; \\      \textcolor{sky_blue}{\textbf{FILTER}} keep filtrate ;\\      \textcolor{sky_blue}{\textbf{CONCENTRATE}} ;\\      \textcolor{sky_blue}{\textbf{YIELD}} \$-1\$ (204 mg).\end{tabular} \\ \bottomrule
    \end{tabular}
    \caption{Example 5.}
\end{subtable}
\begin{subtable}{\textwidth}
\begin{tabular}{p{1.5cm}l|p{1.5cm}l} \toprule
\textbf{Field} & \multicolumn{3}{l}{\textbf{Value}}                          \\ \midrule
REACTANT       & \multicolumn{3}{l}{\$3\$:   Cc1ccc2c(N)ccc(O)c2n1}                    \\
               & \multicolumn{3}{l}{\$4\$: O=N{[}O-{]}.{[}Na+{]}}                      \\
               & \multicolumn{3}{l}{\$5\$:   {[}N-{]}={[}N+{]}={[}N-{]}.{[}Na+{]}}     \\ \midrule
Solvent        & \multicolumn{3}{l}{\$1\$: Cl}                                         \\
               & \multicolumn{3}{l}{\$2\$: O}                                          \\ \midrule
PRODUCT        & \multicolumn{3}{l}{\$-1\$:   Cc1ccc2c(N={[}N+{]}={[}N-{]})ccc(O)c2n1} \\ \midrule
Source &
    \multicolumn{3}{p{13cm}}{5-Amino-8-hydroxy-2-methylquinoline (12; 723 mg, 4.2   mmol) was dissolved in a solution of concentrated hydrochloric acid (0.4 mL)   and water (5 mL), cooled to -3° C. in a salt-ice bath, stirred for 10 min,   then treated dropwise with a cold solution of sodium nitrite (0.50 g, 7.2   mmol) in water (5 mL). The mixture was stirred for 20 min, then treated   dropwise with sodium azide (0.60 g, 9.2 mmol) in water (40 mL), stirred at 0°   C. for a further 1.5 h, then allowed to warm to room temperature over 24 h in   the dark. Isolation by extraction with diethyl ether gave a dark brown solid,   which was recrystallized from light petroleum to yield 13 as light brown   crystals (554 mg, 66\%). 1H NMR (500 MHz, CD3OD) $\delta$ ppm: 3.31 (s, 3H), 7.06 (d,   J=8.5 Hz, 1H), 7.15 (d, J=8.5 Hz, 1H), 7.36 (d, J=8.5 Hz, 1H), 8.21 (d, J=8.5   Hz, 1H). 13C NMR (125 MHz, CD3OD) $\delta$ ppm: 24.9, 111.8, 115.6, 121.6, 123.7,   127.9, 132.6, 139.6, 151.1, 159.7. HRMS (ESI): calcd for C10H8N4O {[}M+H{]}+:   201.0771, found: 201.0773.} \\ \midrule
Annotation   Actions &
\begin{tabular}[c]{@{}p{5cm}@{}}\textcolor{sky_blue}{\textbf{MAKESOLUTION}}   with \$1\$ (0.4 mL) and \$2\$ (5 mL) ; \\      \textcolor{sky_blue}{\textbf{ADD}} \$3\$ (723 mg, 4.2 mmol) ; \\      \textcolor{sky_blue}{\textbf{SETTEMPERATURE}} -3 °C ; \\      \textcolor{sky_blue}{\textbf{STIR}} for 10 min ; \\      \textcolor{sky_blue}{\textbf{MAKESOLUTION}} with \$4\$ (0.50 g, 7.2 mmol) and \$2\$ (5 mL) ; \\      \textcolor{sky_blue}{\textbf{STIR}} for 20 min ; \\      \textcolor{sky_blue}{\textbf{MAKESOLUTION}} with \$5\$ (0.60 g, 9.2 mmol) and \$2\$ (40 mL) ; \\      \textcolor{sky_blue}{\textbf{STIR}} for 1.5 h at 0 °C ; \\      \textcolor{sky_blue}{\textbf{RECRYSTALLIZE}} from petroleum ; \\      \textcolor{sky_blue}{\textbf{YIELD}} \$-1\$ (554 mg, 66\%).\end{tabular} &
    Predicted Actions & 
    \begin{tabular}[c]{@{}p{6cm}@{}}\textcolor{sky_blue}{\textbf{MAKESOLUTION}}   with \$1\$ (3 ml) and \$2\$ (1 ml) ;\\       \textcolor{sky_blue}{\textbf{ADD}} \$3\$ (210 mg, 1.24 mmol) ; \\      \textcolor{sky_blue}{\textbf{STIR}} for 1 h at 0 °C; \\      \textcolor{sky_blue}{\textbf{MAKESOLUTION}} with \$4\$ (80 mg, 1.24 mmol) and \$2\$ (1 ml) ; \\      \textcolor{sky_blue}{\textbf{STIR}} for 1 h at  0 °C; \\      \textcolor{sky_blue}{\textbf{MAKESOLUTION}} with \$5\$ (88 mg, 1.24 mmol) and \$2\$ (1 ml) ; \\      \textcolor{sky_blue}{\textbf{STIR}} for 2 h at 0 °C ; \\      \textcolor{sky_blue}{\textbf{RECRYSTALLIZE}} ; \\      \textcolor{sky_blue}{\textbf{YIELD}} \$-1\$ (120 mg, 47\%).\end{tabular} \\ \bottomrule
\end{tabular}
\caption{Example 6.}
\end{subtable}
\caption{Examples of experimental procedure predictions that are different from the annotation but might be viable.}
\label{tab:case_56}
\end{table*}

\begin{table*}[t]
\small
\centering
\begin{tabular}{p{1.5cm}l|p{1.5cm}l} \toprule
    \textbf{Field} & \multicolumn{3}{l}{\textbf{Value}}                                            \\ \midrule
Reactant       & \multicolumn{3}{l}{\$1\$:  CNC(=O)c1cn(CCCCc2ccc(N)nn2)nn1}                     \\
               & \multicolumn{3}{l}{\$2\$:  O=C(O)Cc1cc(Br)ccn1}                                 \\
               & \multicolumn{3}{l}{\$4\$:    CCCP1(=O)OP(=O)(CCC)OP(=O)(CCC)O1}                 \\
               & \multicolumn{3}{l}{\$6\$:    CCN(C(C)C)C(C)C}                                   \\ \midrule
Solvent        & \multicolumn{3}{l}{\$3\$:  CN(C)C=O}                                            \\
               & \multicolumn{3}{l}{\$5\$: CCOC(C)=O}                                            \\
               & \multicolumn{3}{l}{\$7\$: O}                                                    \\ \midrule
Product        & \multicolumn{3}{l}{\$-1\$:   CNC(=O)c1cn(CCCCc2ccc(NC(=O)Cc3cc(Br)ccn3)nn2)nn1} \\ \midrule
Source &
  \multicolumn{3}{p{13cm}}{To a solution of   1-(4-(6-aminopyridazin-3-yl)butyl)-N-methyl-1H-1,2,3-triazole-4-carboxamide   (2.0 g, 7.2 mmol) and 2-(4-bromopyridin-2-yl)acetic acid (1.6 g, 7.2 mmol) in   DMF (5 ml) were added T3P (50 wt. \% in EtOAc, 9.2 g, 14 mmol) and DIEA (2.7   g, 22 mmol). The mixture was stirred at RT for 3 h, then water (50 mL) was   added, the mixture was stirred for 30 min, and precipitate was isolated by   filtration to give the title compound as a yellow solid (1 g, 40\%). MS (ES+)   C19H21BrN8O2 requires: 473. found: 474 {[}M+H{]}+.} \\ \midrule
Annotated Actions &
  \begin{tabular}[c]{@{}p{5.5cm}@{}} \textcolor{sky_blue}{\textbf{MAKESOLUTION}} with \$1\$ (2.0 g,   7.2 mmol) and \$2\$ (1.6 g, 7.2 mmol) and \$3\$ (5 ml) ;\\      \textcolor{sky_blue}{\textbf{MAKESOLUTION}} with \$4\$ and \$5\$ (9.2 g, 14 mmol) and \$6\$ (2.7 g, 22 mmol)   ;\\      \textcolor{sky_blue}{\textbf{STIR}} for 3 h at RT ;\\      \textcolor{sky_blue}{\textbf{ADD}} \$7\$ (50 mL) ;\\      \textcolor{sky_blue}{\textbf{STIR}} for 30 min ;\\      \textcolor{sky_blue}{\textbf{YIELD}} \$-1\$ (1 g, 40\%).\end{tabular} &
  Predicted Actions &
  \begin{tabular}[c]{@{}p{6cm}@{}}\textcolor{sky_blue}{\textbf{MAKESOLUTION}} with \$1\$ (30 mg,   0.109 mmol) and \$2\$ (23 mg, 0.109 mmol) and \$3\$ (0.5 ml) ; \\      \textcolor{sky_blue}{\textbf{ADD}} \$4\$ (67 l, 0.109 mmol, 0.109 mmol) ;\\      \textcolor{sky_blue}{\textbf{STIR}} for 3 h at room temperature ;\\      \textcolor{sky_blue}{\textbf{ADD}} \$5\$ (5 ml) ;\\      \textcolor{sky_blue}{\textbf{ADD}} \$6\$ (67 l, 0.38 mmol, 0.38 mmol) ;\\      \textcolor{sky_blue}{\textbf{STIR}} for 30 min at room temperature ;\\      \textcolor{sky_blue}{\textbf{ADD}} \$7\$ (0.5 ml) ;\\      \textcolor{sky_blue}{\textbf{CONCENTRATE}} ;\\      \textcolor{sky_blue}{\textbf{YIELD}} \$-1\$ (36 mg, 70\% yield).\end{tabular} \\ \bottomrule
\end{tabular}
\caption{Examples of experimental procedure predictions that are different from the annotation but might be viable. Example 7.}
\label{tab:case_7}
\end{table*}


\vspace{-3mm}
\section{More Experimental Results}
\subsection{Ablation Study}
Table~\ref{tab:ablate_caption} presents an ablation study examining the impact of input contexts on molecule captioning. The removal of the random molecule context results in diminished captioning performance. This observation can be attributed to two factors: 1) including the PubChem324k dataset, which is used for creating random molecule contexts, is important to maintain molecule captioning performance; and 2) without random molecule contexts, the LM becomes overly dependent on reaction contexts, compromising its capability to accurately caption individual molecules. This finding underscores the significance of incorporating random molecule contexts in training.

\subsection{Statistical Analysis}
\label{sec:app-t_test}

\shiyr{We carry out statistical tests on the experimental results to demonstrate that ReactXT achieves a significant performance improvement compared to the baseline models. For most metrics (such as BLEU, ROUGE, METEOR), we employ the T-test; for Top-k accuracy, where calculating the standard deviation was challenging, we use a 2-proportion Z-test instead.}

\shiyr{The results of the statistical tests are presented in Tables \ref{tab:t_test_action} to \ref{tab:t_test_retro}.  We \textbf{bold} p-values that are smaller than 0.05. From these tables, it can be observed that our method achieves statistically significant improvements across all metrics within the tasks of experimental procedure prediction (Table \ref{tab:t_test_action}) and molecule captioning (Tables \ref{tab:t_test_caption_a} and \ref{tab:t_test_caption_b}). As for the retrosynthesis task (Table \ref{tab:t_test_retro}), our method demonstrates statistically significant enhancements in both Top1 and Top3 accuracies. These observations collectively demonstrate the effectiveness of our proposed pretraining method.}

\subsection{Case Studies and Error Analysis}
\subsubsection{Experimental Procedure Prediction}
In this section, we present case studies from the experimental procedure prediction task to inform future research. We include examples of accurate predictions (see Table~\ref{tab:case_12}), inaccurate predictions (see Tables~\ref{tab:case_34}), and predictions that are different from the annotations but may also work (see Table~\ref{tab:case_56} and Table~\ref{tab:case_7}). Our selection criteria prioritizes the accuracy of action sequences and the correct identification of primary materials, while overlooking specifics like material quantities and temperatures. All the examples are from the test set of OpenExp.

Table~\ref{tab:case_12} displays two examples where experimental procedures are accurately predicted, showing close alignment between predicted and annotated actions, albeit with slight variances in material quantities and experiment times.  These cases highlight the capability of LMs to predict experimental procedures, suggesting a path toward automating chemical synthesis.

Table~\ref{tab:case_34} displays two failed examples of experimental procedure prediction. The predicted action sequences significantly deviate from the annotated sequences, making them impractical. Additionally, we can observe one common error of repetition, with the same or similar actions being duplicated.

Tables~\ref{tab:case_56} and Table~\ref{tab:case_7} showcase three examples where the predictions, while different from the annotations, could still be viable. In Example 5, as an alternative to the annotated 'EXTRACT with ethyl acetate', the model proposes a series of actions (`COLLECT LAYER', `WASH with ethyl acetate', `DRY SOLUTION', and `FILTER'), serving a similar function. In Example 6, instead of the specified 'SET TEMPERATURE' and 'STIR', the model recommends `STIR for 1h at 0 °C', serving the same purpose. In Example 7, the model suggests adding components (`ADD \$4\$', `ADD \$5\$', `ADD \$6\$') sequentially rather than making a single solution as annotated, which could also be effective.

\subsubsection{Human Evaluation of OpenExp}
\label{sec:app-human_cases}

In this section, we present case studies from human evaluations on the OpenExp dataset. Samples rated from 5 to 1 by human evaluators are included, as shown in Tables~\ref{tab:openexp_case_5} to Tables~\ref{tab:openexp_case_1}. All samples are from the 250 human evaluated data points (see Appendix~\ref{sec:app-openexp}). It can be observed that samples with two or fewer errors may only have minor flaws, such as typol errors or incorrect numerical values.

\begin{table*}[t]
    \small
    \centering
    \begin{tabular}{p{1.5cm}l} \toprule
        \textbf{Field} & \textbf{Value}                                            \\ \midrule
    Source &
    \multicolumn{1}{p{13cm}}{To a solution of 2-(5-amino-3-methyl-1H-pyrazol-4-yl)-benzothiazole-5-carboxylic acid ethyl ester (30 mg) in THF (1 mL) was added lithium aluminum hydride (4 mg). The reaction mixture was stirred at room temperature for 5 hrs at which point sodium sulfate nonahydrate was added. The resulting mixture was stirred for an additional 30 min. The solids were removed by filtration. The solvent was then evaporated and the residue was purified by flash column chromatography eluting with CHCl3:MeOH=9:1 to yield 21 mg (81\%) of the title compound as a cream coloured solid. MS (m/z, ES+): 261.1 (M+1, 100\%).}
    \\ \midrule
    Annotated Actions &
        \begin{tabular}[c]{@{}p{13cm}@{}}
        \textcolor{sky_blue}{\textbf{MAKESOLUTION}} with 2-(5-amino-3-methyl-1H-pyrazol-4-yl)-benzothiazole-5-carboxylic acid ethyl ester (30  ; mg) and THF (1 mL) ; \\
        \textcolor{sky_blue}{\textbf{ADD}} lithium aluminum hydride (4 mg) ; \\
        \textcolor{sky_blue}{\textbf{STIR}} for 5 hr at room temperature ; \\
        \textcolor{sky_blue}{\textbf{ADD}} sodium sulfate nonahydrate ; \\
        \textcolor{sky_blue}{\textbf{STIR}} for 30 min at room temperature ; \\
        \textcolor{sky_blue}{\textbf{FILTER}} keep filtrate ; \\
        \textcolor{sky_blue}{\textbf{CONCENTRATE}} ; \\
        \textcolor{sky_blue}{\textbf{YIELD}} PRODUCT (21 mg, 81\%) .
        \end{tabular} \\
    \bottomrule
    \end{tabular}
    \caption{Example with a Human Evaluation Score of 5. The action sequence accurately captures the source paragraph.}
    \label{tab:openexp_case_5}
\end{table*}

\begin{table*}[t]
    \small
    \centering
    \begin{tabular}{p{1.5cm}l} \toprule
        \textbf{Field} & \textbf{Value}                                            \\ \midrule
    Source &
    \multicolumn{1}{p{13cm}}{A mixture of (5-nitro-pyridin-2-yl)-(2,2,2-trifluoro-ethyl)-amine (230 mg, 1.04 mmol), cesium carbonate (730 mg, 2.07 mmol) and iodomethane (0.59 mL, 4.18 mmol) in DMF (4 mL) was heated in a sealed tube at 50° C. for 3 hr. The reaction mixture was evaporated to dryness and the crude was partitioned between methylene chloride and water. The organic layer was dried over magnesium sulfate, filtered and concentrated to give methyl-(5-nitro-pyridin-2-yl)-(2,2,2-trifluoro-ethyl)-amine (270 mg, crude) as a brown solid, which was directly used in the next step reaction without further purification. LCMS calcd for C8H8F3N3O2 (m/e) 235, obsd 236 (M+H).}
    \\ \midrule
    Annotated Actions &
        \begin{tabular}[c]{@{}p{13cm}@{}}
        \textcolor{sky_blue}{\textbf{MAKESOLUTION}} with (5-nitro-pyridin-2-yl)-(2,2,2-trifluoro-ethyl)-amine (230 mg, 1.04 mmol) and cesium carbonate (730 mg, 2.07 mmol) and iodomethane (0.59 mL, 4.18 mmol) and DMF (4 mL) ; \\
        \textcolor{sky_blue}{\textbf{STIR}} for 3 hr at 50 °C ; \\
        \textcolor{sky_blue}{\textbf{CONCENTRATE}} ; \\
        \textcolor{sky_blue}{\textbf{PARTITION}} with methylene chloride and water ; \\
        \textcolor{sky_blue}{\textbf{COLLECTLAYER}} organic ; \\
        \textcolor{sky_blue}{\textbf{DRYSOLUTION}} over magnesium sulfate ; \\
        \textcolor{sky_blue}{\textbf{FILTER}} keep \textcolor{mygreen}{filtrate} ; \\
        \textcolor{sky_blue}{\textbf{CONCENTRATE}} ; \\
        \textcolor{sky_blue}{\textbf{YIELD}} PRODUCT (270 mg) .
        \end{tabular} \\
    \bottomrule
    \end{tabular}
    \caption{Example with a Human Evaluation Score of 4. The action sequence contains 1 error, which is highlighted in \textcolor{mygreen}{green}.}
    \label{tab:openexp_case_4}
\end{table*}

\begin{table*}[t]
    \small
    \centering
    \begin{tabular}{p{1.5cm}l} \toprule
        \textbf{Field} & \textbf{Value}                                            \\ \midrule
    Source &
    \multicolumn{1}{p{13cm}}{To a stirred solution of 1,5-anhydro-2,3-dideoxy-D-erythro-hexitol (44.9 g) and imidazole (65.2 g) in DMF (500 ml) was added tert-butylchlorodiphenylsilane (88.5 mL) at 0° C. After stirring for 4 h, the reaction mixture was diluted with EtOAc (1000 ml). The organic layer was washed with water (200 mL×5) and brine (200 mL), and dried over Na2SO4. The solution was concentrated under reduced pressure, and the residue was purified by column chromatography (PE/EtOAc) to afford 70.9 g of the title compound as a colorless oil.}
    \\ \midrule
    Annotated Actions &
        \begin{tabular}[c]{@{}p{13cm}@{}}
        \textcolor{sky_blue}{\textbf{MAKESOLUTION}} with 1,5-anhydro-2,3-dideoxy-D-erythro-hexitol (44.9 g) and imidazole (65.2 g) and DMF (500 ml) ; \\
        \textcolor{sky_blue}{\textbf{ADD}} tert-butylchlorodiphenylsilane (88.5 mL) at 0 °C ; \\
        \textcolor{mygreen}{\textbf{ADD} tert-butylchlorodiphenylsilane (88.5 mL)} ; \\
        \textcolor{sky_blue}{\textbf{STIR}} for 4 h at 0 °C ; \\
        \textcolor{sky_blue}{\textbf{ADD}} ethyl acetate (1000 ml) ; \\
        \textcolor{sky_blue}{\textbf{COLLECTLAYER}} organic ; \\
        \textcolor{sky_blue}{\textbf{WASH}} with water (\textcolor{mygreen}{200 mL}) ; \\
        \textcolor{sky_blue}{\textbf{WASH}} with brine (200 mL) ; \\
        \textcolor{sky_blue}{\textbf{DRYSOLUTION}} over Na2SO4 ; \\
        \textcolor{sky_blue}{\textbf{FILTER}} keep filtrate ; \\
        \textcolor{sky_blue}{\textbf{CONCENTRATE}} ; \\
        \textcolor{sky_blue}{\textbf{YIELD}} PRODUCT (70.9 g) .
        \end{tabular} \\
    \bottomrule
    \end{tabular}
    \caption{Example with a Human Evaluation Score of 3. The action sequence contains 2 errors, which are highlighted in \textcolor{mygreen}{green}.}
    \label{tab:openexp_case_3}
\end{table*}

\begin{table*}[t]
    \small
    \centering
    \begin{tabular}{p{1.5cm}l} \toprule
        \textbf{Field} & \textbf{Value}                                            \\ \midrule
    Source &
    \multicolumn{1}{p{13cm}}{A mixture of methyl 3-hydroxy-1-methyl-1H-pyrazole-5-carboxylate (2.34 g, 15.0 mmol), iodomethane (3.19 g, 22.5 mmol), potassium carbonate (4.15 g, 30.0 mmol) and N,N-dimethylformamide (15 ml) was stirred at room temperature for 18 hr. The mixture was diluted with water (50 mL), and extracted with ethyl acetate (50 mL×3). The organic layer was washed with water (10 mL×2), and concentrated under reduced pressure. The residue was purified by silica gel column chromatography (hexane/ethyl acetate=100/0→50/50) to give the title compound (2.01 g, yield 79\%) as a white solid. 1H-NMR (DMSO-d6, 300 MHz)  3.78 (3H, s), 3.81 (3H, s), 3.94 (3H, s), 6.27 (1H, s).}
    \\ \midrule
    Annotated Actions &
        \begin{tabular}[c]{@{}p{13cm}@{}}
        \textcolor{sky_blue}{\textbf{MAKESOLUTION}} with methyl 3-hydroxy-1-methyl-1H-pyrazole-5-carboxylate (2.34 g, 15.0 mmol) and iodomethane (3.19 g, 22.5 mmol) and potassium carbonate (4.15 g, 30.0 mmol) and N,N-dimethylformamide (15 ml) ; \\
        \textcolor{sky_blue}{\textbf{STIR}} for 18 hr at room temperature ; \\
        \textcolor{sky_blue}{\textbf{ADD}} water (50 mL) at room temperature \textcolor{mygreen}{over 18 hr} ; \\
        \textcolor{mygreen}{\textbf{COLLECTLAYER} organic} ; \\
        \textcolor{mygreen}{\textbf{WASH} with ethyl acetate (50 mL)} ; \\
        \textcolor{sky_blue}{\textbf{COLLECTLAYER}} organic ; \\
        \textcolor{sky_blue}{\textbf{WASH}} with water (\textcolor{mygreen}{10 mL}) ; \\
        \textcolor{sky_blue}{\textbf{CONCENTRATE}} ; \\
        \textcolor{sky_blue}{\textbf{YIELD}} PRODUCT (2.01 g, yield 79\%) .
        \end{tabular} \\
    \bottomrule
    \end{tabular}
    \caption{Example with a Human Evaluation Score of 2. The action sequence contains 4 errors, which are highlighted in \textcolor{mygreen}{green}.}
    \label{tab:openexp_case_2}
\end{table*}

\begin{table*}[t]
    \small
    \centering
    \begin{tabular}{p{1.5cm}l} \toprule
        \textbf{Field} & \textbf{Value}                                            \\ \midrule
    Source &
    \multicolumn{1}{p{13cm}}{4-[4-(4-Fluoro-phenyl)-thiazol-2-yl]-2'-nitro-biphenyl-2-carboxylic acid Sodium hydroxide (40 mg, 1 mmol) was added to a suspension of 4-[4-(4-fluoro-phenyl)-thiazol-2-yl]-2'-nitro-biphenyl-2-carboxylic acid methyl ester (40 mg) in a mixture of water (1 mL) and dioxane (1 mL). The resulting mixture was heated at 50° C. for 4 h. The solvent was evaporated and water (5 mL) was added. The mixture was filtered and the filtrate was made acidic to pH 3 by the addition of concentrated HCl. The precipitate was collected by filtration and dried to give 4-[4-(4-fluoro-phenyl)-thiazol-2-yl]-2'-nitro-biphenyl-2-carboxylic acid (45 mg, 58\% for two steps). 1H NMR (300 MHz, DMSO-d6)  13.21 (br s, 1H), 8.65 (s, 1H), 8.17-8.33 (m, 5H), 7.86 (t, J=7.6 Hz, 1H), 7.73 (t, J=7.8 Hz, 1H), 7.52 (t, J=8.5 Hz, 2H), 7.40 (t, J=8.6 Hz, 2H).}
    \\ \midrule
    Annotated Actions &
        \begin{tabular}[c]{@{}p{13cm}@{}}
        \textcolor{sky_blue}{\textbf{MAKESOLUTION}} with 4-[4-(4-fluoro-phenyl)-thiazol-2-yl]-2'-nitro-biphenyl-2-carboxylic acid methyl ester (40 mg) and \textcolor{mygreen}{4-[4-(4-fluoro-phenyl)-thiazol-2-yl]-2'-nitro-biphenyl-2-carboxylic acid methyl ester (40 mg)} and water (1 mL) and dioxane (1 mL) ; \\
        \textcolor{sky_blue}{\textbf{ADD}} 4-[4-(4-Fluoro-phenyl)-thiazol-2-yl]-2'-nitro-biphenyl-2-carboxylic acid Sodium hydroxide (40 mg, 1 mmol) ; \\
        \textcolor{sky_blue}{\textbf{MAKESOLUTION}} with water (1 mL) and dioxane (1 mL) ; \\
        \textcolor{sky_blue}{\textbf{STIR}} for 4 h at 50 °C ; \\
        \textcolor{sky_blue}{\textbf{CONCENTRATE}} ; \\
        \textcolor{sky_blue}{\textbf{ADD}} water (5 mL) ; \\
        \textcolor{sky_blue}{\textbf{FILTER}} keep \textcolor{mygreen}{precipitate} ; \\
        \textcolor{mygreen}{\{missing operation\}} ; \\
        \textcolor{sky_blue}{\textbf{DRYSOLUTION}} over \textcolor{mygreen}{4-[4-(4-fluoro-phenyl)-thiazol-2-yl]-2'-nitro-biphenyl-2-carboxylic acid (45 mg)} ; \\
        \textcolor{mygreen}{\textbf{FILTER} keep filtrate} ; \\
        \textcolor{sky_blue}{\textbf{YIELD}} PRODUCT (45 mg, 58\%) .
        \end{tabular} \\
    \bottomrule
    \end{tabular}
    \caption{Example with a Human Evaluation Score of 1. The action sequence contains 5 errors, which are highlighted in \textcolor{mygreen}{green}.}
    \label{tab:openexp_case_1}
\end{table*}

\end{document}